\documentclass[twocolumn,amsmath,amssymb,floatfix,prb,showpacs,superscriptaddress]{revtex4}
\usepackage{amsmath,amssymb,natbib,bm,graphicx,times}
\usepackage[hyperfigures,colorlinks=true,allcolors=blue]{hyperref}

\newcommand{\NN}{\mathbb{N}}

\newcommand{\vc}[1]{\mathbf{#1}}

\newcommand{\abs}[1]{\left|#1\right|}
\newcommand{\bra}[1]{\langle \, #1 \,|}
\newcommand{\ket}[1]{|\, #1 \, \rangle}
\newcommand{\bket}[2]{\langle \, #1 \,|\, #2 \, \rangle}
\newcommand{\boket}[3]{\langle\, #1 \,|\, #2 \,|\, #3 \,\rangle}
\newcommand{\ketbra}[1]{\ket{#1} \bra{#1}}

\newcommand{\trr}{$$\to$$}

\newcommand{\be}{\begin{equation}}
\newcommand{\ee}{\end{equation}}
\newcommand{\bc}{\begin{center}}
\newcommand{\ec}{\end{center}}

\begin{document}
\title{Circuit QED with fluxonium qubits: theory of the dispersive regime}
\author{Guanyu Zhu}
\affiliation{Department of Physics and Astronomy, Northwestern University, Evanston, IL 60208, USA}
\author{David G. Ferguson}
\affiliation{Departments of Physics and Astronomy, Northwestern University, Evanston, IL 60208, USA}
\author{Vladimir E. Manucharyan}
\affiliation{Society of Fellows, Harvard University, Cambridge, MA 02138, USA}
\author{Jens Koch}
\affiliation{Departments of Physics and Astronomy, Northwestern University, Evanston, IL 60208, USA}
\date{\today}

\begin{abstract}
In circuit QED, protocols for quantum gates and readout of superconducting qubits often rely on the dispersive regime, reached when the qubit-photon detuning $\Delta$ is large compared to the mutual coupling strength. For qubits including the Cooper-pair box and transmon, selection rules dramatically restrict the contributions to dispersive level shifts $\chi$. 
 By contrast, in the absence of selection rules many virtual transitions contribute to $\chi$ and can produce sizable  dispersive shifts even at large detuning. We present theory for a generic multi-level qudit capacitively coupled to one or multiple harmonic modes,  and give general expressions for the effective Hamiltonian in second and fourth order perturbation theory. Applying our results to the fluxonium  system, we show that the absence of strong selection rules explains the surprisingly large dispersive shifts observed in experiments and  leads to the prediction of a two-photon vacuum Rabi splitting. Quantitative predictions from our theory are in good agreement with experimental data over a wide range of magnetic flux and reveal that fourth-order resonances are important for the phase modulation observed in fluxonium spectroscopy. \end{abstract}
\maketitle

\section{Introduction}
In many respects, the quantum physics of superconducting circuits\cite{makhlin_quantum-state_2001,devoret_implementing_2004,clarke_superconducting_2008} resembles that of atoms: both systems feature a set of discrete, non-equidistant energy levels, and both can be probed by virtue of their interaction with photons. Within circuit QED\cite{blais_cavity_2004,wallraff_strong_2004,schoelkopf_wiring_2008}, this interaction is harnessed to manipulate and measure the quantum state of the superconducting circuit with great success.

In many realizations of the circuit QED architecture, the dispersive regime plays a particularly important role for implementing the readout and gate operations required for a universal quantum computer. The general idea behind the dispersive regime is simple: when detuning the qubit frequency far from the frequency of the resonator, the  interaction-induced conversion of a qubit excitation into a photon becomes ineffective. Specifically, the probability amplitude for the conversion is proportional to the small parameter $g/\Delta$, where $g$ denotes the coupling strength and $\Delta$ the detuning between the qubit and photon frequency. Accordingly, the dispersive regime is the primary setting for performing qubit gates.\cite{blais_cavity_2004,schuster_ac_2005,wallraff_approaching_2005,blais_quantum-information_2007,gywat_dynamics_2006,majer_coupling_2007}

In the dispersive regime, the qubit-photon coupling manifests in the form of energy shifts of Lamb and ac-Stark type.\cite{blais_cavity_2004,schuster_ac_2005,wallraff_approaching_2005,gambetta_qubit-photon_2006,gambetta_protocols_2007,tornberg_dispersive_2007,serban_crossoverweak-_2007,boissonneault_nonlinear_2008,boissonneault_dispersive_2009,zueco_qubit-oscillator_2009,filipp_two-qubit_2009} Here,  ac-Stark shifts correspond to \emph{state-dependent} energy shifts which, in the simplest case, take on the form $\chi a^\dag a\sigma_z$. This expression can be interpreted as a
frequency shift of the resonator, with the size of the shift depending on the state of the qubit, or alternatively, as a shift of the qubit transition frequency, with the size of the shift depending on the photon state of the resonator. Consequently, the dispersive regime provides both a convenient means of qubit readout.\cite{schuster_ac_2005,wallraff_approaching_2005,gambetta_qubit-photon_2006,Siddiqi2006,gambetta_protocols_2007,tornberg_dispersive_2007,serban_crossoverweak-_2007,blais_quantum-information_2007,filipp_two-qubit_2009,Reed2010,Wirth2010}, as well as a measurement tool for investigating the resonator state.\cite{gambetta_qubit-photon_2006,schuster_resolving_2007} Possible limitations to this simple picture due to corrections of higher order in the parameter $g/\Delta$ have recently been studied by Boissonneault et al.\cite{boissonneault_nonlinear_2008,boissonneault_dispersive_2009}

The physics of the dispersive regime becomes richer when higher levels of the superconducting circuit (which we hence refer to as \emph{qudit}) participate in the virtual transitions that contribute to the dispersive shifts. The simplest manifestation of this is the contribution of the third level of the transmon to the dispersive shift, even under conditions when real occupation of this level is negligible. For specific level configurations relative to the resonator frequency, the two partial contributions to $\chi$ add up constructively and give rise to the straddling regime with characteristically large dispersive shifts.\cite{koch_charge-insensitive_2007,Boissonneault2012} Recently, enhanced dispersive shifts have also been predicted and confirmed experimentally for a flux qubit coupled to a resonator.\cite{Inomata2012} Similar to the transmon straddling regime, higher levels of the flux qubit are responsible for the observed shift enhancement.

\begin{figure}[b]
\includegraphics[width=1.0\columnwidth]{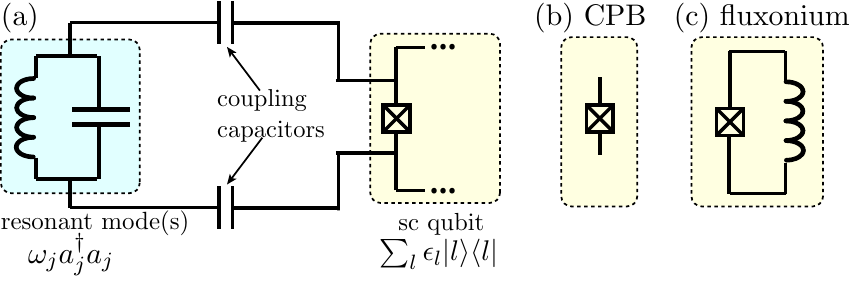}
\caption{(Color online) (a) General scheme of circuit quantum electrodynamics with capacitive coupling between one (or several) resonant mode(s) and a superconducting circuit acting as a qudit. Examples of superconducting qudits are atoms are (b) the Cooper pair box in the charging and transmon regime, and (c) the fluxonium circuit. \label{fig:circuitQED}}
\end{figure}

In this paper, we present theory that systematically describes the dispersive regime for a generic circuit QED system consisting of a multi-level qudit coupled capacitively to one or multiple harmonic modes.  In Section \ref{sect-general}, we derive the general expression of the effective Hamiltonian governing the dispersive regime up to (and including) terms of fourth order in the coupling between the qudit and the harmonic modes. We verify that our results correctly reproduce the well-known expressions for the dispersive regime of the Cooper pair box in the charging and transmon regime. In Section \ref{fluxoniumcase} we then apply our results to the  fluxonium system\cite{manucharyan_fluxonium:_2009,koch_charging_2009,phase_slip_2012} where, different from the Cooper pair box case, the lack of selection rules allows for a large number of terms to contribute to the dispersive shifts.
We compare our theoretical predictions with the data from reflection and spectroscopy experiments obtained previously by the Yale group,\cite{manucharyan_fluxonium:_2009,phase_slip_2012, Vlad-thesis} and summarize our findings and conclusions in Section \ref{conclusions}.

\section{Dispersive regime of charge-coupled circuit QED systems\label{sect-general}}
\subsection{General Model}
We consider a charge-coupled circuit QED system\cite{blais_cavity_2004,wallraff_strong_2004,schoelkopf_wiring_2008} as depicted in Fig.~\ref{fig:circuitQED}. The system contains a superconducting circuit\cite{makhlin_quantum-state_2001,devoret_superconducting_2003,clarke_superconducting_2008} acting as a quantum system with a discrete and anharmonic energy spectrum. Honoring the multi-level nature, we will refer to it as a qudit and denote its eigenstates and corresponding energies  as $\ket{l}$ and $\epsilon_l$, respectively. Here, $l$ enumerates the eigenstates starting with $l=0$ for the qudit ground state. The qudit is coupled to microwave photons inside one or several\cite{johnson_quantum_2010}
 superconducting transmission line resonator(s) with normal mode frequencies $\omega_j$. We consider a finite number of such modes enumerated by $j=1,2,\ldots$ The inclusion of the infinite set of higher modes inside each resonator is beyond the scope of the present paper, but has been subject of recent studies\cite{Filipp2011,blais-unpublished} indicating that the qubit size provides a natural cutoff in $j$. 

The generic model Hamiltonian $H= H_0+ V$ describing the coupled circuit QED system thus captures the bare qudit and harmonic modes,
\begin{align}
H_0=&\sum_{j}\omega_j a^\dag_j a^{\phantom{\dag}}_j + \sum_l \epsilon_l \ket{l}\bra{l}
\label{Hamiltonian1}
\end{align}
and the interaction between them,
\begin{align}
 {V}=&\sum_j\sum_{l,l'} g_{j;ll'} \ket{l}\bra{l'}(a_j+a_j^\dag).
\label{Hamiltonian2}
\end{align}
Here, $a_j$ ($a_j^\dag$) is the usual annihilation (creation) operator for a photon in mode $j$. 
The interaction term describes the coupling between the relevant charge variable of the qudit, $(2e)\mathsf{N}$, and the electric voltage of the resonator at the qudit position, $V_j=V^\text{rms}_j(a_j^\dag+a_j)$. The resulting coupling coefficients are given by
\be
g_{j;ll'} = g_j \boket{l}{ \mathsf{N}}{l'},
\ee
where $g_j= 2e \beta_j V^\text{rms}_j$ abbreviates the qudit-independent part of the coupling strength and $\beta_j$ is a dimensionless capacitance ratio, typically of order unity.\cite{koch_charge-insensitive_2007} We set $\hbar=1$ and, in the following, only deviate from this convention when discussing concrete experimental parameters and observables.

The above Hamiltonian provides a generic model for simple, charge-coupled circuit QED systems. Adopting the appropriate qudit energy spectrum $\{\epsilon_l\}$ and  coupling parameters $g_{j;ll'}$, Equations \eqref{Hamiltonian1} and \eqref{Hamiltonian2} may model, for instance, circuit QED systems based on the Cooper pair box, the transmon,\cite{koch_charge-insensitive_2007,schreier_suppressing_2008} or the recently developed fluxonium device.\cite{manucharyan_fluxonium:_2009} All three examples are summarized in Table \ref{table1}. Note that the interaction, in general, does \emph{not} conserve the overall excitation number $\sum_j a_j^\dag a_j+\sum_l l\ketbra{l}$ unless special selection rules restrict the coupling $g_{j;ll'}$ to nearest-neighbor qudit levels and rotating wave approximation (RWA) is assumed.

\begin{table}[b]
\centering
\caption{Level structure and selection rules for three different superconducting (sc) qubits, capacitively coupled to a single resonator mode. The frequency $\omega_p=\sqrt{8E_JE_C}$ denotes the plasma oscillation frequency, and $E_J$ and $E_C$ are the Josephson and charging energy, respectively. For the Cooper pair box in both the charging ($E_J\ll E_C$) and transmon $(E_J\gg E_C)$ regime, simplifying selection rules apply. No simple selection rules exist for the fluxonium device. (Abbreviations used: TLS -- two-level system, MLS -- multi-level system.)\label{table1}}
\begin{ruledtabular}
\begin{tabular}{lll}
\textbf{sc qubit} & \textbf{level structure} & \textbf{selection rules} \\\hline\hline
CPB/charge & $\epsilon_0=0,\,\epsilon_{1}$ & $g_{10}=g_{01}=g$\\
& (treated as TLS) & other $g_{ll'}=0$ \\\hline
CPB/transmon & $\epsilon_{l}\simeq l\,\omega_p$ & $g_{l,l\pm1}$\\
& $\qquad-E_C(l^2+l)/2$ &other $g_{ll'}\simeq0$ \\
& (weakly anharmonic MLS) \\\hline
fluxonium & $\epsilon_{l}$ & no simple \\
& (anharmonic MLS) & selection rules
\end{tabular}
\end{ruledtabular}
\end{table}

\subsection{Effective Hamiltonian for the Dispersive Regime}
The qudit-photon interaction $V$ facilitates transitions $\ket{l}\to\ket{l'}$ between qudit states which are accompanied by the emission or absorption of photons. The dispersive regime of circuit QED\cite{blais_cavity_2004,boissonneault_nonlinear_2008,boissonneault_dispersive_2009,zueco_qubit-oscillator_2009,Ong2011} describes the situation when such transitions are suppressed due to large detuning between the qudit transition frequencies and the relevant  mode frequencies. The adiabatic elimination of the coupling $V$, appropriate when the energy mismatch is large compared to the coupling strength, has found many applications in different branches of physics and, depending on context, is known under several names including van-Vleck perturbation theory\cite{vanVleck_1926} and Schrieffer-Wolff transformation.\cite{schrieffer_1966}

Adiabatic elimination is illustrated most simply for an energy spectrum featuring a large gap between two groups of unperturbed states. By construction of an appropriate canonical transformation, one obtains an effective Hamiltonian in which the weak interaction of states above the gap with states below, is eliminated in favor of dressed states with energies slightly shifted relative to the unperturbed ones. The setting of two subspaces separated by a single large gap, however, is not a necessary requirement for the approach. The typical situation of the dispersive limit in circuit QED indeed differs from that simple setting:  each unperturbed state $\ket{\vc{n}l}_0$  forms its individual subspace as long as transitions from one qudit state to another remain sufficiently detuned from the photon frequencies. Here, stats are labeled by the set of photon numbers $\vc{n}=(n_1,n_2,\ldots)$ and qudit states $l=0,1,\;ldots$ The contribution of \emph{several} virtual transitions to each energy level shift can make the physics of the dispersive regime quite rich. In part, this is already true for transmon-based circuit QED systems, where the $l=2$ state gives rise to enhanced level shifts in the straddling regime.\cite{koch_charge-insensitive_2007} Even more interesting features emerge in the dispersive regime of the fluxonium device, which we discuss in detail in Section \ref{fluxoniumcase}. Based on the lucid description given by Cohen-Tannoudji et al.,\cite{cohen-tannoudji_atom-photon_1998}
we next summarize the systematic procedure for obtaining the effective dispersive Hamiltonian of the circuit QED model specified in Eqs.\ \eqref{Hamiltonian1} and \eqref{Hamiltonian2}.

 As a necessary condition for the validity of the dispersive approximation, all one-photon transitions among low-lying qudit states must be strongly detuned from the harmonic mode frequencies. To formulate this condition quantitatively, we introduce compact notation for transition energies and detunings: $\epsilon_{ll'}\equiv\epsilon_{l}-\epsilon_{l'}$  abbreviates the energy released in the qudit transition $l\trr l'$  (note that $\epsilon_{ll'}$ is negative when $l'>l$), and $\Delta_{j;ll'} \equiv\epsilon_{ll'}-\omega_j$  denotes the detuning between this transition and resonator mode $j$. In this notation, the condition for the dispersive regime reads
\be\label{dispersivecondition}
|\Delta_{j;ll'}|\gg |g_{j;ll'}|\sqrt{n_j+1},
\ee
where the  photon number is typically restricted to $n_j=0$ when assuming dilution refrigerator temperatures $k_{\!B}T\ll\omega_j$, but may reach higher values when the system is driven with microwave tones.

Condition \eqref{dispersivecondition} motivates the perturbative treatment of the interaction $V$ which couples the unperturbed $H_0$ eigenstates $\ket{\vc{n}l}_0$.\footnote{Note that the subscript ``0'' for bare states is suppressed in the text up to Eq.\ \eqref{dispersivecondition}, but is stated explicitly from here on to distinguish bare from dressed states.} Each eigenstate of $H$ is a dressed state with the majority of all probability amplitudes in a single state $\ket{\vc{n}l}_0$. As a result, the  labeling of bare states can be maintained for the dressed eigenstates  $\ket{\vc{n}l}=e^{-iS}\ket{\vc{n}l}_0$. The diagonalization of $H$, up to a specified order in $V$, is achieved by the unitary transformation ${H}'=e^{i {S}} H e^{-i {S}}$.  Note that, by construction, eigenstates of $H'$ are just the unperturbed states $\ket{\vc{n}l}_0$. 

The procedure\cite{cohen-tannoudji_atom-photon_1998} for obtaining the required hermitean generator $S$ now follows from  the two conditions that $H'$ be diagonal and the generator $S$ be off-diagonal in the unperturbed basis:
\be\label{diagonalcondition}
P_{\vc{n}l}\,H'P_{\vc{n}'l'}\sim\delta_{\vc{n}\vc{n}'}\delta_{ll'} \quad\text{and}\quad
P_{\vc{n}l}\,SP_{\vc{n}l}=0.
\ee
Here, $P_{\vc{n}l}\equiv\ket{\vc{n}l}_0{_0\bra{\vc{n}l}}$ is the projector onto a single unperturbed state. Introducing an auxiliary parameter $\lambda$ for counting powers in $V$, one constructs  
$S= \sum_{m=1}^{\infty}{\lambda}^m  {S}_m$ and $ H'={H}_0+ \sum_{m=1}^{\infty}{\lambda}^m {H}'_m$ order by order, by comparing with the nested commutator series
\be
   {H}'=e^{i {S}} H e^{-i {S}}=\sum_{m=0}^{\infty}\frac{1}{m!}[i {S}, {H}]_m
\ee
and enforcing the conditions \eqref{diagonalcondition}.  (For further details, see Appendix  \ref{adiabatic}.)

In practice, this procedure quickly becomes cumbersome for terms beyond second order. Since terms  of fourth order in the interaction will turn out to be relevant for the dispersive regime of fluxonium, we devise a way to bypass the evaluation of fourth-order nested commutators as follows. Note that there is a natural equivalence between the construction of the generator $S$ on one hand, and the  ordinary form of time-independent perturbation theory (yielding corrections to energies and states) on the other hand. The two approaches merely differ in whether the basis change to the approximate eigenbasis is carried out as an active or passive transformation. In the first approach, the perturbation determines the generator $S$ which brings the Hamiltonian $H$ into the diagonal form of $H'$. In the second approach, the perturbation affects the dressed states which one  constructs explicitly in the unperturbed basis as $\ket{\vc{n}l}=e^{-iS}\ket{\vc{n}l}_0$. Fortunately, obtaining higher-order corrections for eigenenergies in the second approach generally does not involve nested commutators. Thus,  we first establish the generic structure of $H'$ up to the desired order, leaving all energy coefficients of individual terms to be determined. We then apply the inverse unitary transformation (cut off at the same order) and obtain the effective Hamiltonian $H_\text{eff}=e^{-iS}H'e^{iS}$. Finally, we employ ordinary perturbation theory to find the eigenenergy corrections and extract from them the undetermined energy coefficients to complete the effective Hamiltonian. 

The generic form of $H'$ is dictated by the conditions from Eq.\ \eqref{diagonalcondition}, and is easily obtained  as follows. Since Eq.\ \eqref{diagonalcondition} excludes all coupling between different subspaces, $H'$ can be expressed as 
\begin{align}
 {H}'=&\sum_{\vc{n},l} {P}_{\vc{n}l} {H}' {P}_{\vc{n}l}= \sum_{\vc{n},l} E_{\vc{n}l}{P}_{\vc{n}l}.
\end{align}
Any operator contributing to $H'=\sum_kH_k'$  must be diagonal in the unperturbed basis, i.e., ${P}_{\vc{n}l} {H_k'}  {P}_{\vc{n}l}  \neq 0$.
Evidently, each such contribution can only consist of harmonic-mode number operators and qudit projectors. The resulting general form, after performing the inverse unitary transformation,  is
\be\label{hkprime}
H_{\text{eff};k}=\alpha_k \prod_j (\mathsf{a}_j^\dag \mathsf{a}_j)^{N_{jk}}\ketbra{l_k}.
\ee
Here $N_{jk}\ge 0$ are integer exponents, $\alpha_k$ is an energy coefficient, and the harmonic oscillator and projection operators are dressed-state operators, i.e.,
\be
\mathsf{a}_j=e^{iS}a_j\,e^{-iS} \quad  \text{and} \quad \ketbra{l_k}=e^{iS}\ket{l_k}_0{_0\bra{l_k}}e^{-iS}.
\ee
Since the interaction ${V}$ is of order one and consists of a sum over operator terms with only one harmonic ladder operator each, the perturbation order $\lambda^m$ of any contribution~\eqref{hkprime} cannot be smaller than the number of ladder operators, i.e, $m\ge 2\sum_j N_{jk}$. This provides us with the necessary information to obtain the generic structure of the effective Hamiltonian.

\begin{figure}
\centering
\includegraphics[width=1.0\columnwidth]{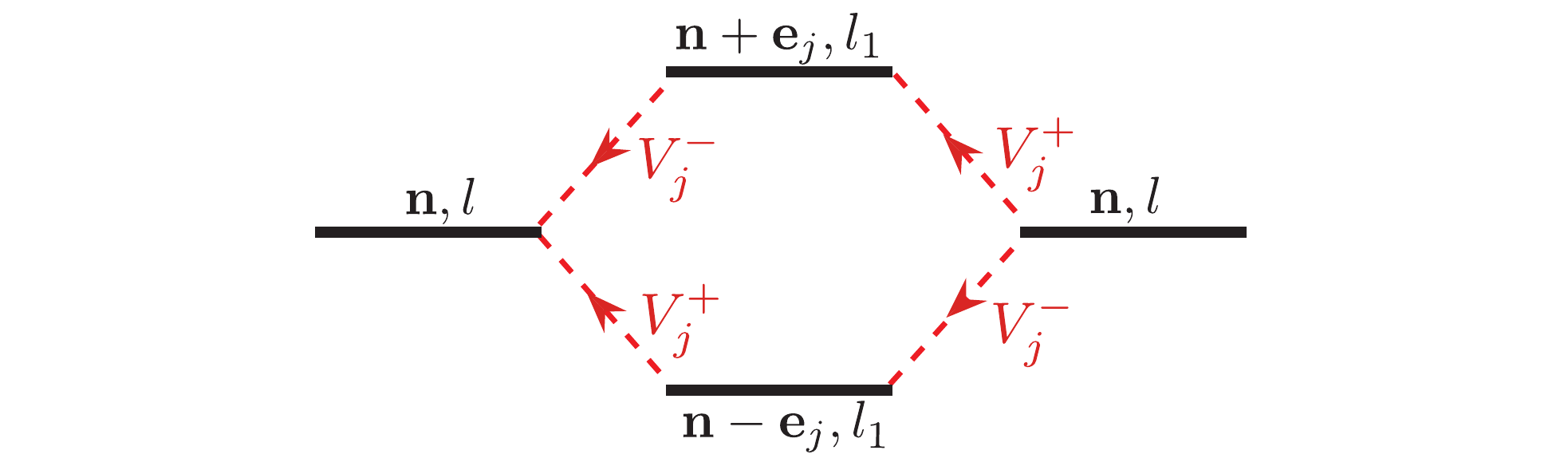}
\caption{(Color online)  Second-order ladder diagram showing the contribution of harmonic mode $j$ to the eigenenergy correction for state $\ket{\vc{n},l}$. The two interfering paths correspond to virtual transitions which intermediately decrease ($V_j^+V_j^-$) and increase ($V_j^-V_j^+$) the photon number in mode $j$. Without selection rules,  summation includes all qudit levels $l_1$. $\vc{e}_j$ denotes the unit vector with ``direction'' $j$.  \label{fig-ladder1}}
\end{figure}

\begin{figure}
\includegraphics[width=1.0\columnwidth]{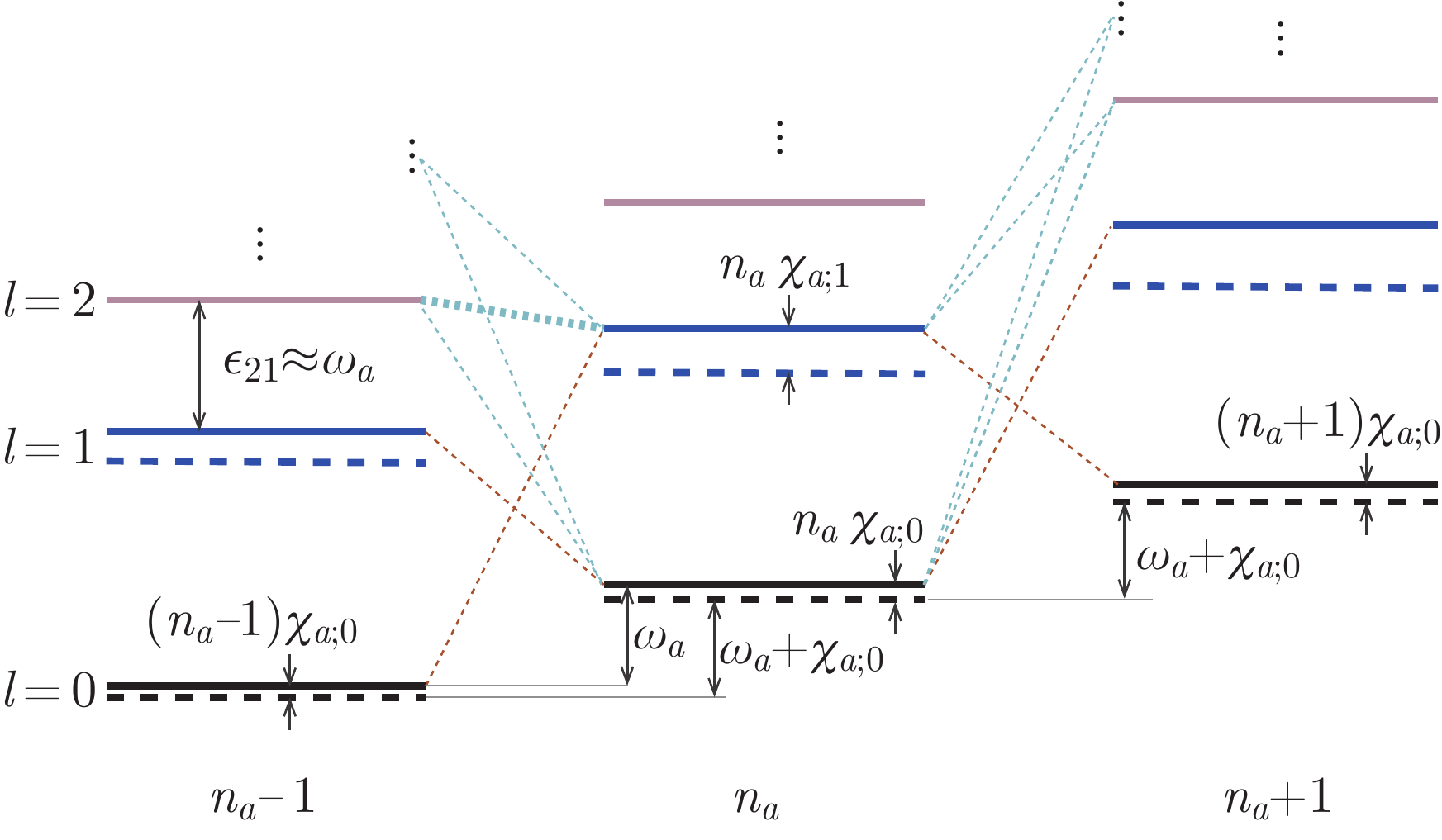}
\caption{(Color online)  Level diagram illustrating the second-order dispersive shifts for a multi-level qudit coupled to a single photon mode.  For each column (definite photon number),  solid lines show the bare energy levels with $l=0, 1 \ \text{and}\ 2$ qudit excitations.  Diagonal dashed lines show the relevant couplings between the $l=0, 1$ levels with $n_a$ photons (middle column) to states with $n_a\pm1$ photons  (red lines: coupling $l=0,1$ states,  blue lines: coupling to higher states). Note that  higher levels (e.g.\ $l=2$) participate in the couplings and affect the dispersive shifts.  The shifted levels for the $l=0,1$ states are shown as horizontal dashed lines.  In this particular example, the qudit transition $\epsilon_{21}$ is nearly resonant with the photon frequency $\omega_a$ leading to a large shift of the $l=1$ level. 
\label{haroche}}
\end{figure}

\emph{Second-order terms.}---The generic structure of the effective Hamiltonian in second-order perturbation theory is
\begin{align}\label{second}
\nonumber H_{\text{eff}}=& \sum_{j}\omega_j \mathsf{a}^\dag_j \mathsf{a}_j + \sum_l \epsilon_l \ket{l}\bra{l} \\
& +\sum_{j,l}  \chi_{j;l} \mathsf{a}_{j}^{\dag} \mathsf{a}_j \ketbra{l} +\sum_l\kappa_l\ketbra{l}.
\end{align}
Here, the third and fourth terms describe dispersive shifts of ac-Stark type and qudit level shifts of Lamb type. As usual, the ac-Stark shifts may manifest as harmonic-mode frequency shifts which depend on the occupied qudit level $l$, $\omega_{j} \to \omega_j+\chi_{j;l}$, or as  qudit energy shifts which depend on photon numbers,  $\epsilon_{l}\to\epsilon_l + \sum_j \chi_{j;l}\mathsf{a}^\dag_j \mathsf{a}_j$.\cite{gambetta_qubit-photon_2006,schuster_resolving_2007}
Note that terms of the form $\Delta\omega_j \mathsf{a}_{j}^{\dag} \mathsf{a}_j$  corresponding to pure shifts of resonant mode frequencies may be absorbed by letting $\chi_{j;l}\to\chi_{j;l}+\Delta\omega_j$. To determine the coefficients $\chi_{j;l}$ and $\kappa_l$ we use the ordinary expression for the second-order energy correction:
\begin{align}\label{secondorderenergy}
E_{\vc{n}l}^{(2)}
= {\sum_{\vc{n}'l'}}' \frac{{_0\boket{\vc{n}l}{V}{\vc{n}'l'}}_0{_0\boket{\vc{n}'l'}{V}{\vc{n}l}}_0}{E_{\vc{n}l}^{(0)}-E_{\vc{n}'l'}^{(0)}},
\end{align}
where the primed sum indicates that the term $\vc{n}'l'=\vc{n}l$ is to be omitted. For further evaluation, we separate the interaction into photon creation and annihilation terms of the individual modes, $V=\sum_j(V_j^++V_j^-)$ where 
\begin{align}\label{vpm}
{V}_j^+ = \sum_{l,l'}g_{j;l'l}\mathsf{a}_j^\dag \ket{l'}\bra{l} \quad \text{and} \quad V_j^-=(V_j^+)^\dag.
\end{align}
The product of transition matrix elements in the numerator of Eq.\ \eqref{secondorderenergy} selects the combinations $V_j^+V_j^-$ and $V_j^-V_j^+$, which change the photon number by one and subsequently undo this change. The virtual transitions affecting photon number and qudit levels are illustrated conveniently in the ladder diagram shown in Fig.\ \ref{fig-ladder1}. They yield
\[\textstyle
E_{\vc{n}l}^{(2)}=\sum_{j,l'}(n_j[\chi_{j;ll'}-\chi_{j;l'l}]+\chi_{j;ll'})
\]
for the second-order energy correction, where 
\be
\chi_{j;ll'} \equiv {\smash |g_{j;ll'}}|^2/\Delta_{j;ll'}
\ee
 abbreviates partial dispersive shifts. The wanted energy coefficients in Eq.~\eqref{second} can now be read off. The resulting expressions are given by
\begin{align}\label{coeffs1}
\chi_{j;l}
= \sum_{l'}(\chi_{j;ll'} - \chi_{j;l'l}) \quad  \text{and} \quad
\kappa_l  
= \sum_{j,l'} \chi_{j;ll'}.
\end{align}
Note that both expressions include a summation over all qudit levels $l'$. Thus, higher qudit levels -- even when unoccupied -- may contribute substantially to the dispersive shifts of photon frequencies and lower qudit levels. This fact is also clearly illustrated with the level diagram shown in Fig.\ \ref{haroche}.

We have verified that the same expressions are obtained with the active transformation method (see Appendix \ref{adiabatic}). In addition, 
one can readily confirm that Eqs.~\eqref{second} and \eqref{coeffs1} correctly reproduce the following well-known results. First, for the Jaynes-Cummings model with a two-level system (TLS) coupled to a single resonator mode
we have $g_{1;ll'}=g$ for $(l,l')=(1,0)$ or $(0,1)$. For all other choices, $g_{1;ll'}$ vanishes. We thus obtain
\begin{align}
H_\text{eff; JC}
=& \omega \mathsf{a}^\dag \mathsf{a} + \frac{\epsilon_{10}+\chi_{01}}{2}\sigma_z +\chi_{01}\mathsf{a}^\dag \mathsf{a}\sigma_z+\text{const.}
\end{align}
which agrees with the known result.\cite{blais_cavity_2004} Second, we consider the multi-level transmon device coupled to one resonator mode.\cite{koch_charge-insensitive_2007} In this case, there is a selection rule allowing only for coupling of nearest-neighbor transmon levels, i.e. $g_{1;ll'}=0$ for $l'\not=l\pm1$. With this, we recover
\begin{align}
&H_\text{eff; transmon}=\omega \mathsf{a}^\dag \mathsf{a} + \sum_l \epsilon_l \ket{l}\bra{l}\\\nonumber
&\qquad+ \sum_{l>0} (\chi_{l,l-1}-\chi_{l+1,l})\mathsf{a}^\dag \mathsf{a}\ket{l}\bra{l} +  \sum_{l>0} \chi_{l,l-1}\ket{l}\bra{l},
\end{align}
which correctly leads to the expression $(\chi_{01}-\chi_{12}/2)\mathsf{a}^\dag \mathsf{a}\sigma_z$ for the ac-Stark term upon projection onto the subspace spanned by the $l=0$ and $1$ qudit states.\cite{koch_charge-insensitive_2007}

\emph{Fourth-order terms.}---The fourth-order terms of the Hamiltonian take the form
\begin{align}\label{gen4}
\nonumber
&H_{\text{eff}}=[\text{Eq}.\ \eqref{second}]+\sum_{j,l} \chi'_{j;l} \mathsf{a}_j^\dag \mathsf{a}_j \ketbra{l} +\sum_l  \kappa'_l\ketbra{l}\\
&+ \sum_{j,l}\eta_{j;l} (\mathsf{a}_j^\dag \mathsf{a}_j)^2 \ketbra{l} 
 + \sum_{i\not=j,l}\xi_{ij;l} \mathsf{a}_i^\dag \mathsf{a}_i \mathsf{a}_j^\dag \mathsf{a}_j \ketbra{l}.
\end{align}
Beyond additional corrections to terms already present in second order [first line of Eq.\ \eqref{gen4}], the fourth-order terms also introduce interaction among harmonic modes of self-Kerr and cross-Kerr type. We denote the corresponding coefficients by $\eta_{j;l}$ and $\xi_{ij;l}$, respectively. For simplicity (and also motivated by the experimental data to be discussed in Section \ref{fluxoniumcase}), we  focus our discussion on a set of two harmonic modes and refer to them as $\mathsf{a}$ and $\mathsf{b}$ mode.

The fourth-order corrections to the eigenenergies are given by\cite{fourthorder}
\begin{align}\label{4thorderenergy}
  E^{(4)}_\textsc{n} ={ \sum_{\textsc{m},\textsc{p},\textsc{q}}}' \frac{V_\textsc{nm}V_\textsc{mp}V_\textsc{pq}V_\textsc{qn}}{E_\textsc{nm} E_\textsc{np} E_\textsc{nq}} 
 - { \sum_\textsc{m}}' \frac{\abs{V_\textsc{nm}}^2}{E_\textsc{nm}^2}
 { \sum_\textsc{p}}' \frac{\abs{V_\textsc{np}}^2}{E_\textsc{np}} .
\end{align}
Here, $\textsc{n}$, $\textsc{m}$, $\textsc{p}$ and $\textsc{q}$ are multi-indices of the form $\{\vc{n},l\}$. Energy differences in the denominators are given by $E_{\textsc{nm}}\equiv E^{(0)}_\textsc{n} - E^{(0)}_\textsc{m}$. Matrix elements are abbreviated by $V_\textsc{nn'}\equiv \langle \textsc{n} | V | \textsc{n}'\rangle$.
Terms involving diagonal matrix elements have  been dropped  in Eq.\ \eqref{4thorderenergy} since  $V_\textsc{nn}=0$ in the cases of interest. We next sketch the evaluation of the fourth-order corrections and provide the relevant ladder diagrams.

We denote the two terms on the right-hand side of Eq.\ \eqref{4thorderenergy} by (\textsc{i}) and (\textsc{ii}) and start with discussing the latter. Term (\textsc{ii}) is the product of two factors with a structure nearly identical to the  second-order expression \eqref{secondorderenergy}  and thus evaluates to
\begin{align}\nonumber
  E^{(4)\textsc{(ii)}}_{\vc{n}l}=-&\sum_{j,l_1}{\smash |g_{j;ll_1}}|^2\bigg(n_j[\Delta_{j;ll_1}^{-2}-\Delta_{j;l_1l}^{-2}]+\Delta_{j;ll_1}^{-2}\bigg)\\
\times&\sum_{j',l_2}(n_{j'}[\chi_{j';ll_2}-\chi_{j';l_2l}]+\chi_{j';ll_2}).
\end{align}
By expanding this expression, we identify contributions to each of the fourth-order terms given in Eq.\ \eqref{gen4}. In addition to the simple poles like $(\epsilon_{ll'}-\omega_j)^{-1}$ which already appear in the second-order energy coefficients, new double and triple poles with the same denominators $(\epsilon_{ll'}-\omega_j)$ emerge. 

Next, we turn to term (\textsc{i}) in Eq.\ \eqref{4thorderenergy} which cannot be factorized. We classify the contributions from (\textsc{i}) in terms of ladder diagrams. The rules governing these ladder diagrams are as follows:
\begin{enumerate}
\item Each ladder step is labeled by a multi-index $(\vc{n}_m,l_m)$ specifying the occupation of the harmonic modes and the qudit level.
\item Starting from the right, each subsequent ladder step is related to the previous one by a virtual transition effected by the operator $V_j^\pm$ (label on the arrow).
\item The set of all possible paths is constrained by the condition that the left-most state must coincide with the right-most state $(\vc{n},l)$. Thus, each path must contain as many $V_j^+$ as $V_j^-$.
\item Each path traversing from right to left gives a contribution involving summation over all intermediate qudit levels $l_1,l_2,l_3$.
\item The product of matrix elements in the numerator of each term is determined by the sequence of arrow labels in each path. For example, the sequence $V_1^-V_2^- V_2^+ V_1^+$ results in the product $g_{1;ll_3}g_{2;l_3l_2}g_{2;l_2l_1}g_{1;ll_1}$.
\item The denominator of each term consists of a product of three energy differences of the form $E^{(0)}_{\vc{n}l}-E^{(0)}_{\vc{n}'l'}$ where $(\vc{n}',l')$ labels the virtual intermediate states on the inner three ladder steps.
\end{enumerate}
The extraction of the various energy coefficients can further be simplified by  noting that the ladder diagrams also directly specify the operator structure of the resulting  terms in the effective Hamiltonian. For example, the numerator $V^-_1V^+_2V^-_2V^+_1$ is associated with terms of the structure $\mathsf{a}_1\mathsf{a}_2^\dag \mathsf{a}_2 \mathsf{a}_1^\dag=( \mathsf{a}_1^\dag  \mathsf{a}_1+1) \mathsf{a}_2^\dag  \mathsf{a}_2$.

\begin{figure}
\includegraphics[width=1.0\columnwidth]{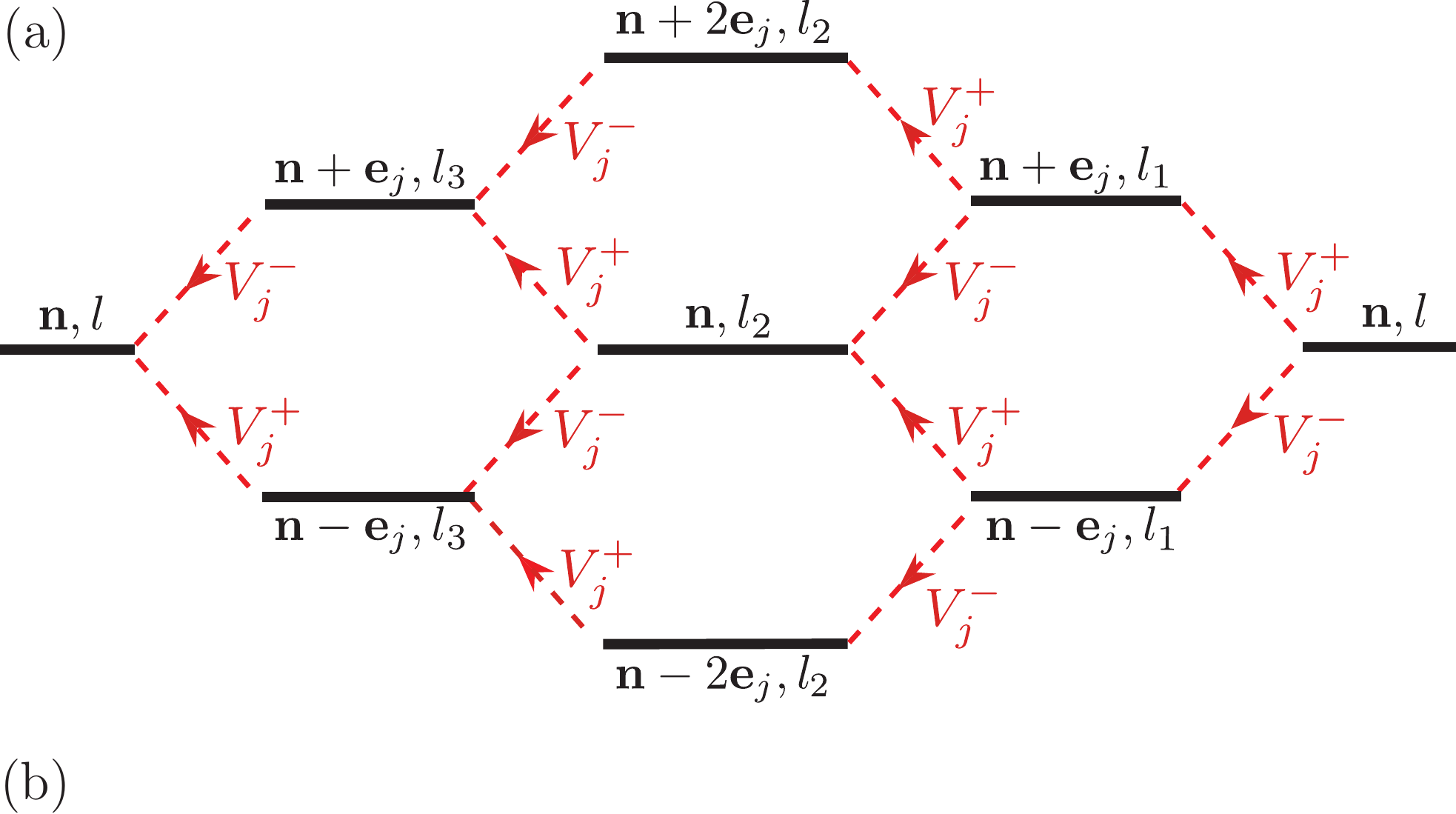}\\[-0.4cm]
{\footnotesize
\begin{tabular}{l|l|l|l|l}
                                           & $E_3$ & $E_2$ & $E_1$ & $\mathsf{op}$ \\\hline\hline
 $V_j^+V_j^+V_j^-V_j^-$ & $\epsilon_{ll_3}+\omega_j$ & $\epsilon_{ll_2}+2\omega_j$ & $\epsilon_{ll_1}+\omega_j$ & $\mathsf{n}_j(\mathsf{n}_j-1)$\\
 $V_j^+V_j^-V_j^+V_j^-$ & $\epsilon_{ll_3}+\omega_j$ & $(\epsilon_{ll_2})'$ & $\epsilon_{ll_1}+\omega_j$ & $(\mathsf{n}_j)^2$\\
 $V_j^+V_j^-V_j^-V_j^+$ & $\epsilon_{ll_3}+\omega_j$ & $(\epsilon_{ll_2})'$ & $\epsilon_{ll_1}-\omega_j$ & $\mathsf{n}_j(\mathsf{n}_j+1)$\\
 $V_j^-V_j^-V_j^+V_j^+$ & $\epsilon_{ll_3}-\omega_j$ & $\epsilon_{ll_2}-2\omega_j$ & $\epsilon_{ll_1}-\omega_j$ & $(\mathsf{n}_j+2)(\mathsf{n}_j+1)$\\
 $V_j^+V_j^-V_j^+V_j^-$ & $\epsilon_{ll_3}-\omega_j$ & $(\epsilon_{ll_2})'$ & $\epsilon_{ll_1}+\omega_j$ & $(\mathsf{n}_j+1)^2$\\
 $V_j^+V_j^-V_j^-V_j^+$ & $\epsilon_{ll_3}-\omega_j$ & $(\epsilon_{ll_2})'$ & $\epsilon_{ll_1}-\omega_j$ & $\mathsf{n}_j(\mathsf{n}_j+1)$\\
\end{tabular}}
\caption{(Color online)  (a) Single-mode ladder diagram  involving only virtual excitations of mode $j$. (b) The corresponding table specifies terms according to Eq.\ \eqref{ladder1}. An additional prime on $E_2$ signals exclusion of the term $l_2=l$ from the sum over $l_2$.   \label{fig3}}
\end{figure}

We first treat the single-mode contributions shown in Fig.\ \ref{fig3}(a). For each given mode $j$, the diagram gives rise to six different paths resulting in terms of the form
\be\label{ladder1}
\sum_l {\sum_{l_1,l_2,l_3}} \frac{g_{j;ll_3}g_{j;l_3l_2}g_{j;l_2l_1}g_{j;l_1l}}{E_3E_2E_1} \mathsf{op}(\mathsf{a}_j,\mathsf{a}_j^\dag)\ketbra{l}.
\ee
The corresponding energy denominators and operators are specified in the table accompanying Fig.\ \ref{fig3}. The first and fourth terms in Fig.\ \ref{fig3}(b) exhibit new poles  absent in second-order perturbation theory. In frequency space, these poles occur when the conditions $\epsilon_{ll_2}\pm 2\omega_j=0$ are met and thus signal additional resonances when qudit transitions match the energy of \emph{two} photons in mode $j$. We will argue below that such resonances have indeed been observed in previous experiments.\cite{manucharyan_fluxonium:_2009,phase_slip_2012,Vlad-thesis}

\begin{figure}
\includegraphics[width=1.0\columnwidth]{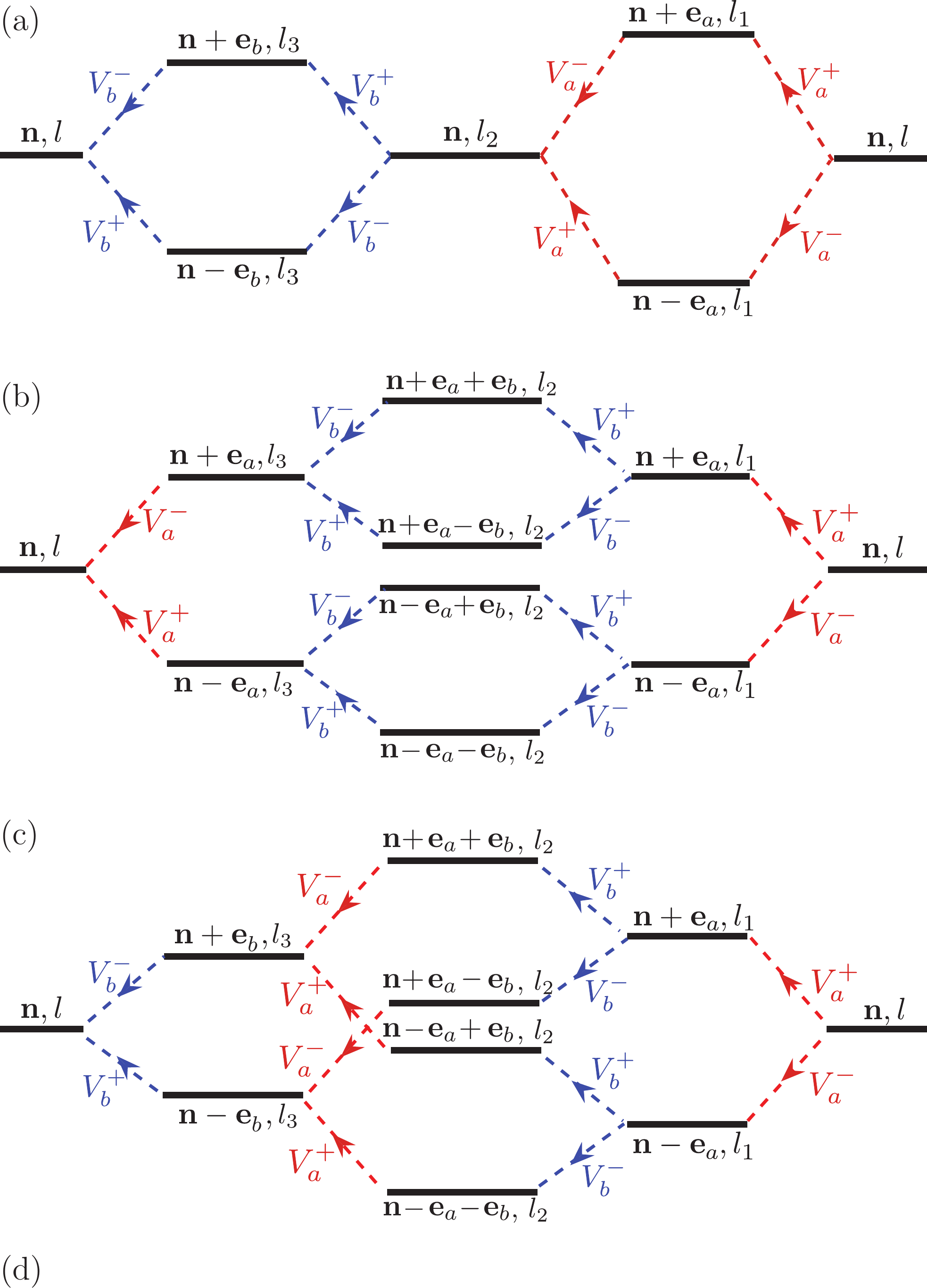}\\[-0.4cm]
{\footnotesize
\begin{tabular}{l|l|l|l|l}
                                           & $E_3$ & $E_2$ & $E_1$ & $\mathsf{op}$ \\\hline\hline
 $V_b^-V_b^+V_a^-V_a^+$ & $\epsilon_{ll_3}-\omega_b$ & $(\epsilon_{ll_2})'$ & $\epsilon_{ll_1}-\omega_a$ & $(\mathsf{n}_a+1)(\mathsf{n}_b+1)$\\
$V_b^+V_b^-V_a^+V_a^-$ & $\epsilon_{ll_3}+\omega_b$ & $(\epsilon_{ll_2})'$ & $\epsilon_{ll_1}-\omega_a$ & $(\mathsf{n}_a+1)\mathsf{n}_b$\\
$V_b^-V_b^+V_a^+V_a^-$ & $\epsilon_{ll_3}-\omega_b$ & $(\epsilon_{ll_2})'$ & $\epsilon_{ll_1}+\omega_a$ & $\mathsf{n}_a(\mathsf{n}_b+1)$\\
$V_b^+V_b^-V_a^+V_a^-$ & $\epsilon_{ll_3}+\omega_b$ & $(\epsilon_{ll_2})'$ & $\epsilon_{ll_1}+\omega_a$ & $\mathsf{n}_a \mathsf{n}_b$\\\hline
 $V_a^-V_b^-V_b^+V_a^+$ & $\epsilon_{ll_3}-\omega_a$  & $\epsilon_{ll_2}-\omega_a-\omega_b$ & $\epsilon_{ll_1}-\omega_a$ & $(\mathsf{n}_a+1)(\mathsf{n}_b+1)$\\
 $V_a^-V_b^+V_b^-V_a^+$ & $\epsilon_{ll_3}-\omega_a$  & $\epsilon_{ll_2}-\omega_a+\omega_b$ & $\epsilon_{ll_1}-\omega_a$ & $(\mathsf{n}_a+1)\mathsf{n}_b$\\
 $V_a^+V_b^-V_b^+V_a^-$ & $\epsilon_{ll_3}+\omega_a$ & $\epsilon_{ll_2}+\omega_a-\omega_b$ & $\epsilon_{ll_1}+\omega_a$ & $\mathsf{n}_a(\mathsf{n}_b+1)$\\
 $V_a^+V_b^+V_b^-V_a^-$ & $\epsilon_{ll_3}+\omega_a$ & $\epsilon_{ll_2}+\omega_a+\omega_b$ & $\epsilon_{ll_1}+\omega_a$ & $\mathsf{n}_a \mathsf{n}_b$\\\hline
 $V_b^-V_a^-V_b^+V_a^+$ & $\epsilon_{ll_3}-\omega_b$ & $\epsilon_{ll_2}-\omega_a-\omega_b$ & $\epsilon_{ll_1}-\omega_a$ & $(\mathsf{n}_a+1)(\mathsf{n}_b+1)$\\
 $V_b^+V_a^-V_b^-V_a^+$ & $\epsilon_{ll_3}+\omega_b$ & $\epsilon_{ll_2}-\omega_a+\omega_b$ & $\epsilon_{ll_1}-\omega_a$ & $(\mathsf{n}_a+1)\mathsf{n}_b$\\
 $V_b^-V_a^+V_b^+V_a^-$ & $\epsilon_{ll_3}-\omega_b$ & $\epsilon_{ll_2}+\omega_a-\omega_b$ & $\epsilon_{ll_1}+\omega_a$ & $\mathsf{n}_a(\mathsf{n}_b+1)$\\
 $V_b^+V_a^+V_b^-V_a^-$ & $\epsilon_{ll_3}+\omega_b$ & $\epsilon_{ll_2}+\omega_a+\omega_b$ & $\epsilon_{ll_1}+\omega_a$ & $\mathsf{n}_a\mathsf{n}_b$
\end{tabular}}
\caption{(Color online)  (a)--(c) Dual-mode ladder diagrams contributing to fourth-order eigenenergy corrections. For a given selection of two modes, labeled $a$ and $b$, each diagram has an associated diagram obtained by interchanging the roles of $a$ and $b$. Only for (a) one finds that label interchange ($a\leftrightarrow b$) yields an identical expression for the resulting contribution. (d) Table detailing the resulting contributions according to Eq.\ \eqref{dual-mode}.\label{fig456}}
\end{figure}

With this motivation in mind, we turn to the remaining fourth-order contributions shown in Fig.\ \ref{fig456}. All of them are of dual-mode type, i.e., they include participation of two different harmonic modes in the virtual transitions.  For clarity, we label the two modes $j=a$ and $j'=b\not=a$ in the following. All resulting contributions can be cast into the form
\be\label{dual-mode}
\sum_l {\sum_{l_1,l_2,l_3}} \frac{g_4g_3g_2g_1}{E_3E_2E_1} \mathsf{op}(\mathsf{a},\mathsf{a}^\dag,\mathsf{b},\mathsf{b}^\dag)\ketbra{l}.
\ee
where $g_\nu=g_{j_\nu;l_\nu l_{\nu-1}}$ with $j_\nu\in\{a,b\}$ and $l_4\equiv l_0\equiv l$. The harmonic lowering operators for the two modes are now simply denoted by $\mathsf{a}$, $\mathsf{b}$. All the possible paths are listed in Fig.\ \ref{fig456}, and the corresponding coefficients and operators are summarized in Fig.\ \ref{fig456}(d).  

New types of poles arise in the ladder diagrams of Fig. \ref{fig456}(b) and (c) when one of the resonance conditions 
\[ \epsilon_{ll_2} \pm (\omega_a+\omega_b)=0 \quad\text{or} \quad \epsilon_{ll_2} \pm (\omega_b- \omega_a)=0\]
 is met. Such resonances occur whenever a qudit transition matches either the energy required for placing one photon each in  mode $a$ and mode $b$, or the energy required to convert an $a$ photon into a $b$ photon. We find that these additional resonances lead to observable effects and can be pinpointed in the data from previous experiments,\cite{manucharyan_fluxonium:_2009,phase_slip_2012,Vlad-thesis} as we will discuss in the following section.

\section{Application: dispersive regime of the fluxonium device\label{fluxoniumcase}}
Equipped with the general expressions for the effective Hamiltonian, we now study their concrete application to the fluxonium circuit.\cite{manucharyan_fluxonium:_2009} The fluxonium device is of particular interest  in this context since experiments have shown surprisingly large dispersive shifts\cite{phase_slip_2012} and, as we will see, transitions between fluxonium states are not strongly restricted by selection rules. The simplest model of fluxonium is obtained by shunting a small Josephson junction with the large kinetic inductance from a Josephson junction array [see Fig.\ \ref{fig:circuitQED}(c)]. For fluxonium, the flux-dependent eigenenergies $\epsilon_l(\Phi_\text{ext})$ and corresponding eigenstates $\{\ket{l}\}$ are thus determined by the Hamiltonian
\be\label{Hf}
H_\text{f} = 4E_C  {\mathsf{N}}^2 -E_J \cos(\varphi-2\pi\Phi_{\text{ext}}/\Phi_0)+\frac{1}{2}E_L{\varphi}^2.
\ee
Here, the operator $ \mathsf{N}= {Q}/2e$ describes the charge on the junction capacitance (in units of the Cooper pair charge) . Its conjugate operator $ {\varphi}=2\pi {\Phi}/\Phi_0$ describes the loop flux in such units that $\varphi$ coincides with the phase difference across the junction. Circuit quantization imposes the commutation relation $[ {\varphi}, \mathsf{N}]= i$ between the two operators. Analogous to the usual relation between position and momentum operators, the charge operator $ \mathsf{N}$ thus takes the form $ \mathsf{N}=-i\frac{d}{d\varphi}$ in $\varphi$ representation. The Hamiltonian $H_\text{f}$ is accompanied by the usual boundary condition $\lim_{\varphi \rightarrow \pm \infty} \psi_l(\varphi) =0$. 

While analytical expressions can be obtained in the limit of very small inductive energies $E_L$,\cite{koch_charging_2009} numerical diagonalization remains the most useful approach for the intermediate parameter values realized in the experiments.\cite{manucharyan_fluxonium:_2009, phase_slip_2012, exp_parameters} Employing the harmonic oscillator basis (which diagonalizes $H_\text{f}$ for $E_J=0$) and taking the necessary precautions for convergence with respect to the basis truncation, we obtain the energy spectrum and eigenstates and calculate the charge matrix elements $\boket{l}{ \mathsf{N}}{l'}$ entering the coupling parameters $g_{j;ll'}$. For the interested reader, we provide details of the numerical diagonalization scheme in Appendix \ref{app:diag}.

Figure \ref{chargematrix} shows representative results for the charge matrix elements obtained in this way, using model parameters which match the experimental values.\cite{exp_parameters} For the example of magnetic flux $\Phi_{\text{ext}}=0.4 \Phi_0$, Fig.\ \ref{chargematrix}(a) shows a broad distribution for $\boket{l}{ \mathsf{N}}{l'}$m  indicating that no strict selection rules apply for the off-diagonal charge matrix elements in the case of fluxonium.  Only at special points where the external flux in units of the flux quantum, $\Phi_{\text{ext}}/\Phi_0$, is integer or half-integer, we recover an odd-even selection rule  caused by the reflection symmetry of the potential with respect to $\varphi=0$. Away from these special points, reflection symmetry is broken and strict selection rules disappear.  Nonetheless, there is clear evidence that certain matrix elements are about an order of magnitude larger than others.  The underlying regularity can be explained by  quasi-selection rules that can be derived analytically \footnote{Guanyu Zhu et al. (in preparation).}.

\begin{figure}
\includegraphics[width=0.9\columnwidth]{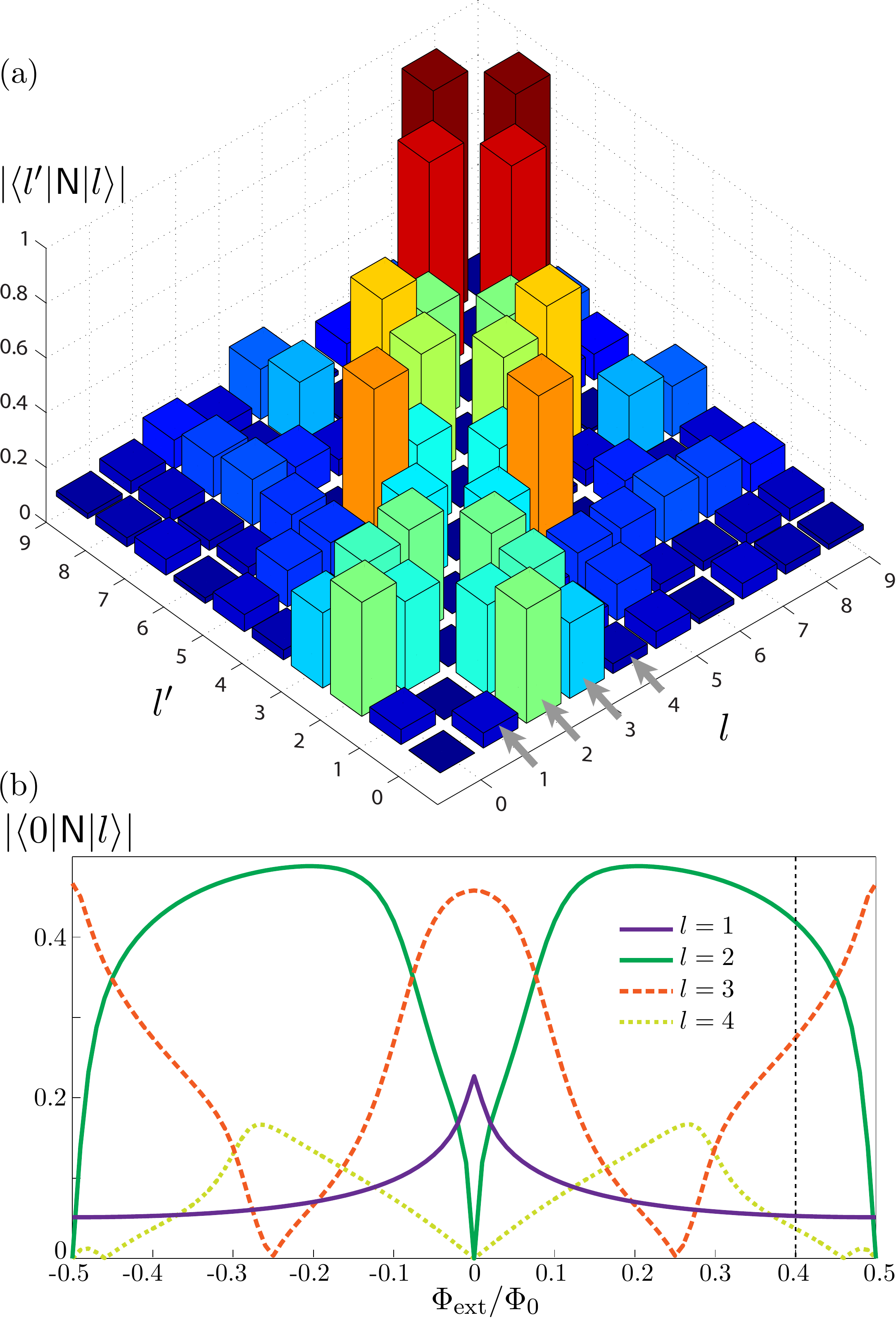}
\caption{(Color online)  (a) Magnitude of the charge matrix elements $|\boket{l}{\mathsf{N}}{l'}|$ between fluxonium states $l$ and $l'$, for external magnetic flux $\Phi_{\text{ext}}=0.4 \Phi_0$. The parameters used match the experimental valuest.\cite{exp_parameters}  Note the absence of a nearest-neighbor selection rule. 
(b) The magnitude of the charge matrix elements between ground state and low-lying excited states, plotted as a function of external magnetic flux. An even/odd selection rule holds for zero and half-integer flux quantum due to parity. Away from these special flux values, simple selection rules are absent. The vertical line marks the flux value $0.4 \Phi_0$ used in (a).}
\label{chargematrix}
\end{figure}

In circuit QED experiments,\cite{manucharyan_fluxonium:_2009, phase_slip_2012} the fluxonium circuit is coupled capacitively to a microwave resonator. To account for an additional harmonic mode observed in the experiment \footnote{We speculate that this mode in fact represents one of the difference modes of the Josephson junction array, see D.~G.~Ferguson et al., arXiv:1208.5747 (2012).} we include coupling to two relevant harmonic modes within a generalized JC Hamiltonian. The first, associated with raising and lowering operators $\mathsf{a}^\dag,\mathsf{a}$,  describes the fundamental mode of the resonator, with frequency $\omega_a$. In the experiment, the resonator supports quarter-wavelength modes only. The lowest harmonic thus has a much higher frequency of $3\omega_a$, and we will neglect the corrections due to higher harmonics in the following. The second mode  represents an observed array mode with frequency $\omega_b$ and coupling strength $g_b$. Both parameters are obtained from a fit to the spectrum in Fig.\ \ref{artificialspectrum}(a).\cite{exp_parameters}

The generic form of the effective Hamiltonian derived in the previous section [see Eq.\ \eqref{gen4}] now takes the concrete form
\begin{align}\label{4thordereffective}
 &H_\text{eff}=\omega_a\mathsf{a}^\dag \mathsf{a} + \omega_b \mathsf{b}^\dag \mathsf{b} + \sum_l(\epsilon_l +\kappa_l+\kappa_l')\ketbra{l}\\
\nonumber & +\sum_l \ketbra{l} \bigg[ (\chi_{a;l}+\chi'_{a;l}) \mathsf{a}^\dag \mathsf{a}  + (\chi_{b;l}+\chi'_{b;l}) \mathsf{b}^\dag \mathsf{b}\\
\nonumber &\qquad\qquad\quad+\eta_{a;l} (\mathsf{a}^\dag \mathsf{a})^2 + \eta_{b;l} (\mathsf{b}^\dag \mathsf{b})^2  + \xi_{ab;l} \mathsf{a}^\dag \mathsf{a} \mathsf{b}^\dag \mathsf{b}\bigg],
\end{align}
where the energies of fluxonium levels as well as their coupling strengths are tunable by the external magnetic flux, $\epsilon_l=\epsilon_l(\Phi_\text{ext})$ and $g_{j;ll'}=g_{j;ll'}(\Phi_\text{ext})$. The ordinary ac-Stark shifts $\chi_{a;l}$  given in Eq.\ \eqref{coeffs1} are now acquire fourth-order contributions. Fourth-order terms are also responsible for the interaction terms given by $\eta_{a;l} (\mathsf{a}^\dag \mathsf{a})^2 \ketbra{l}$ (self-Kerr) and $\xi_{ab;l} \mathsf{a}^\dag \mathsf{a} \mathsf{b}^\dag \mathsf{b}\ketbra{l}$ (cross-Kerr). These terms induce additional nonlinearity and result in dependence of the photon frequencies on the occupation numbers $n_a$ and $n_b$. The effective Hamiltonian for the dispersive regime of the fluxonium circuit  enables us to study the two central types of measurements  performed in Refs.~\onlinecite{manucharyan_fluxonium:_2009,phase_slip_2012}, and \onlinecite{Vlad-thesis}. 

\subsection{Measurements in the dispersive regime\label{sec:measurements}}
The first measurement type is the direct homodyne detection of the reflected amplitude of a single microwave tone. The frequency of this tone is fixed close to the bare resonator frequency, and the voltage of the reflected signal is recorded as a function of magnetic flux. This measurement  primarily probes the dispersive shift of the resonator frequency (i.e., the $\mathsf{a}$ mode)  given by $\chi_{a;l}$ to second order, and by
\begin{align}\label{nonlinearshifts}
\mu_{a;l}(n_a, n_b) \equiv\,&E_{n_a+1,n_b;l}-E_{n_a,n_b;l}-\omega_a\\\nonumber
  =&\chi_{a;l} + \chi'_{a;l}+(2n_a+1)\eta_{a;l}+n_b \xi_{ab; l}
\end{align} 
when including fourth order corrections. Here, the energies $E_{n_a,n_b;l}$ are obtained as the eigenvalues of the effective Hamiltonian, Eq.~\eqref{4thordereffective}. We assume that the fluxonium circuit occupies a fixed level $l$, where usually the ground state $l=0$ is the appropriate state maintained during the measurement. Note that the Kerr terms in fourth order lead to an additional dependence of dispersive shifts on the excitations numbers $n_a$ and $n_b$.

Two-tone spectroscopy is the second measurement type and probes the fluxonium transition frequencies via the $l$ dependence of the resonator shift. The first tone with frequency $\omega_{d1}$ close to the bare resonator frequency acts in a way similar to the one used in the first measurement type. (As an alternative to the reflected amplitude, the phase shift $\theta$ of the reflected tone may be recorded.) A second drive tone is applied and its frequency $\omega_{d2}$ varied over a wide range with the goal of inducing Rabi oscillations between the ground state and an excited state of the fluxonium circuit. The transfer of probability amplitude to  higher levels is accompanied by a change in the dispersive shift of the resonator. In the simplest case, spectroscopy thus probes the corresponding change in the dispersive shift given by $\chi_{a;l} - \chi_{a;0}$ and  $\mu_{a;l} - \mu_{a;0}$ in second and fourth order, respectively.

The detected phase shift $\theta_l$ has a characteristic dependence on the detuning between drive frequency $\omega_d$ and the effective resonator frequency $\omega_a+\chi_{a;l}$ (or $\mu_{a;l}$). To leading order, it follows the characteristic form\cite{phase_slip_2012} 
\be\label{arctanfunction}
 \theta_l \approx 2 \ \text{arctan}\bigg[ \frac{2 Q}{\omega_a}(\omega_d-\omega_a-\chi_{a;l})\bigg]. 
 \ee
Here, $Q$ is the quality factor of the resonator and $\chi$ may be replaced by $\mu$ when considering fourth-order corrections.  
 
In our subsequent discussion of  direct homodyne detection and spectroscopy, the pole structure of the shifts $\chi_{a;l}$ and $\mu_{a;l}$ will play a crucial role. Poles are associated with resonances between fluxonium transitions and harmonic mode excitations, and signal the breakdown of perturbation theory within some frequency window centered at the pole. Since fluxonium levels are tuned by varying the external magnetic flux $\Phi_\text{ext}$, frequency windows will correspond to flux windows in the experiments to be discussed next.

\begin{figure}
\includegraphics[width=1.0\columnwidth]{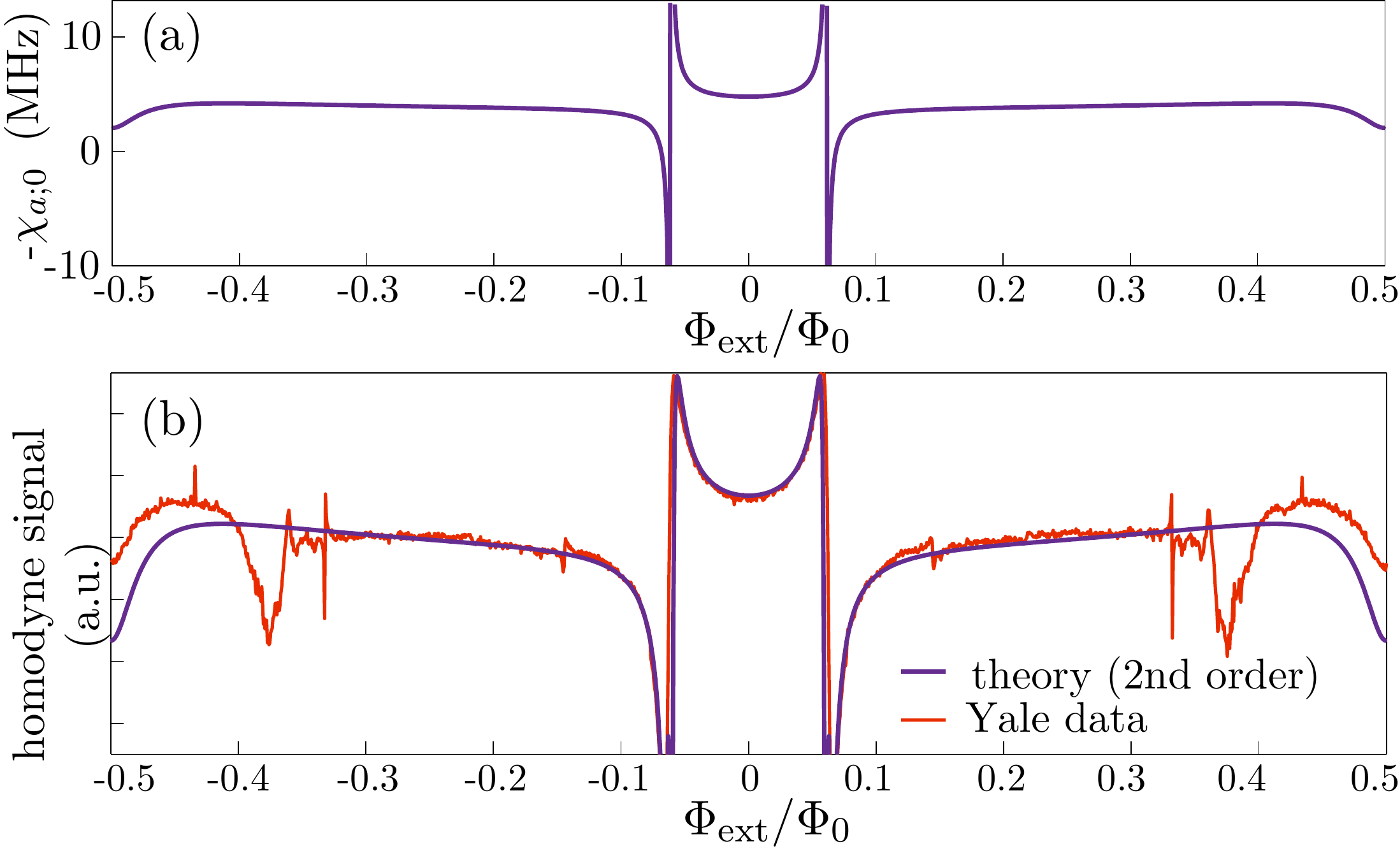}
\caption{(Color online)  (a) Dispersive shift $\chi_{a;0}$ of the resonator frequency for fluxonium ground state, obtained by second-order perturbation theory. For convenient comparison with panel (b), the negative shift $-\chi_{a;0}$ is shown. (b) Comparison between experimental data for direct homodyne detection of the reflected amplitude (red/light gray curve; data as published in Ref.~\onlinecite{manucharyan_fluxonium:_2009}) with theory fit according to Eq.\ \eqref{homodyne_fit} (blue/dark gray curve). The fit parameters are determined as: $ A=0.0094, B=0.0887\text{GHz}^{-1},C=10.3052,\theta_0=-4.543$.
\label{dispersive_shifts}}
\end{figure}

\subsection{Direct homodyne detection of reflected signal}
The specific quantity observed in direct homodyne detection is the quadrature voltage of the reflected signal.\cite{Vlad-thesis} This voltage can be expressed in terms of the phase shift via 
\begin{align}\label{homodyne_fit}
V=A \cos(\theta+\theta'_0)+C
\approx A \cos(B\chi_{a;l}+\theta_0)+C.
\end{align}
Here,  $\theta_0$ and $\theta_0'$ are offset phase shifts which may be present in experimental data, $A$ is the amplitude of the reflected signal and $C$ a constant voltage offset. The approximation in the second step of Eq.~\eqref{homodyne_fit} is obtained by Taylor expansion of Eq.~\eqref{arctanfunction} for the drive close to resonance, i.e., $\abs{\omega_d-\omega_a-\chi_{a;0}}\ll\omega_a$. 

We begin our comparison of theory with the experimental data from Ref.~\onlinecite{manucharyan_fluxonium:_2009} with the expressions obtained in second-order perturbation theory. The predicted dispersive shift of the resonator $\chi_{a;0}$ (while maintaining the $l=0$ ground state) is shown in Fig.~\ref{dispersive_shifts}(a). Its magnitude is of order $1$ to $10\,\text{MHz}$ except in the immediate vicinity of the poles where the fluxonium $0$-$1$ transition crosses the resonator frequency (at flux values $\Phi_\text{ext}/\Phi_0\approx\pm0.06$).

Using Eq.~\eqref{homodyne_fit}, the dispersive shift is converted to homodyne voltage. We adjust the parameters $A$, $B$, $C$ and $\theta_0$ to minimize the mean-square deviations over the magnetic flux range $\Phi_{\text{ext}}/\Phi_0 \in [-0.3,0]$, again assuming occupation of the fluxonium ground state only. We compare the resulting fit with the experimental data in  Fig.\ \ref{dispersive_shifts}(b) and find good agreement in the mentioned flux range with the exception of the small peak-dip structure at $\Phi_\text{ext}/\Phi_0\approx \pm 0.15$. Fourth-order corrections considered below will account for this feature. More significant deviations occur in the flux ranges close to half-integer $\Phi_\text{ext}/\Phi_0$. As we will see, fourth-order corrections from the effective Hamiltonian \eqref{4thordereffective} alone do not lead to a satisfactory resolution, and we will discuss possible culprits for the persistence of deviations in this region. 

\begin{figure}
\includegraphics[width=1.0\columnwidth]{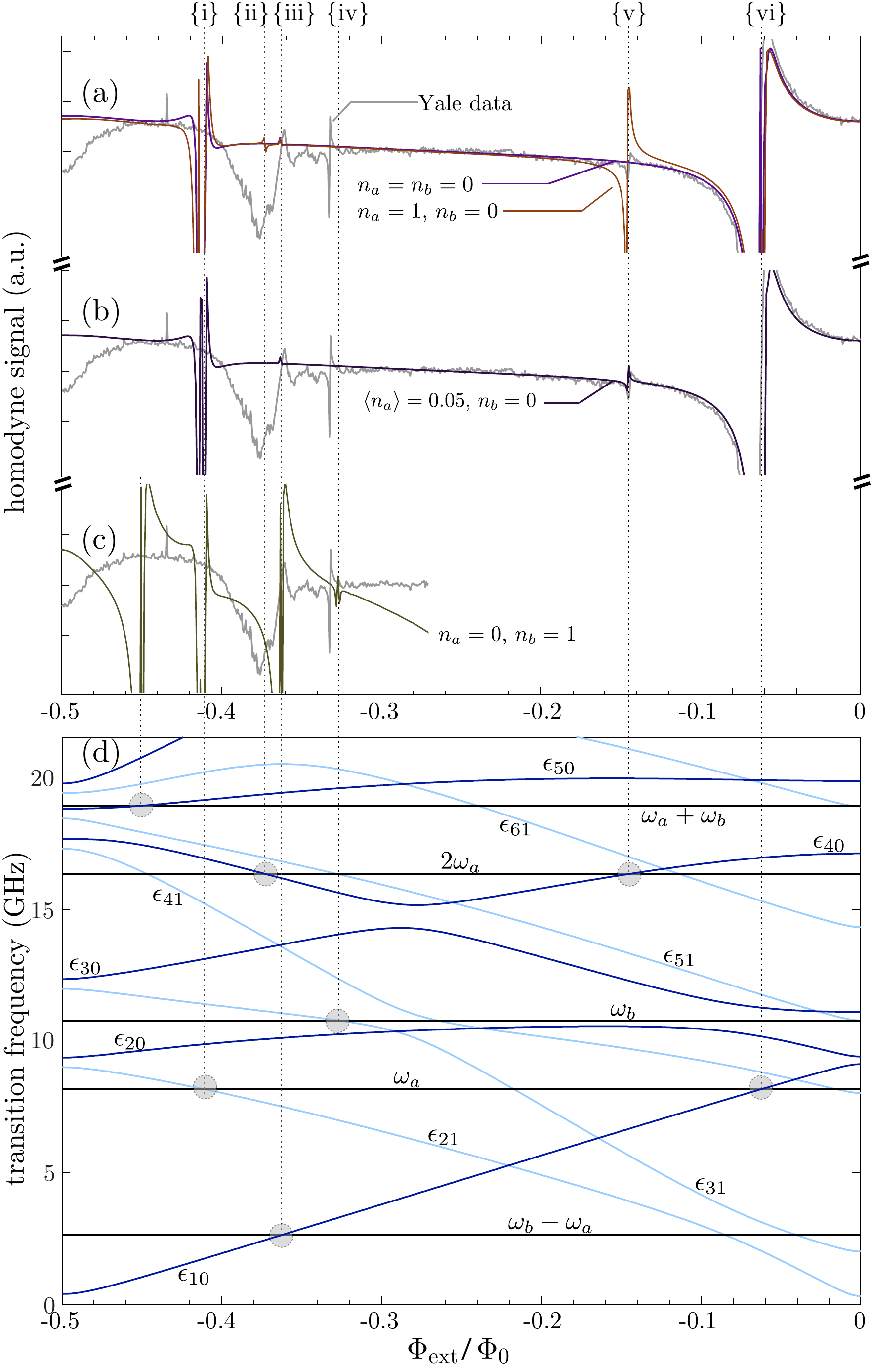}
\caption{(Color online)  Homodyne voltage of a reflected microwave tone: comparison between experimental data from Ref.~\cite{manucharyan_fluxonium:_2009} and fourth-order theory for dispersive shifts, thermal averaging over fluxonium levels according to Eq.~\eqref{Boltzman} included ($T=20\,\text{mK}$).  (a)  Theory curves show dispersive shifts of the resonator frequency  for occupation numbers $n_a\!=\!n_b\!=\!0$ and $n_a\!=\!1,\,n_b\!=\!0$ as indicated. The position of the additional pole at  $\Phi_\text{ext}/\Phi_0\approx \pm 0.15$ is in good agreement with that of the peak-dip structure observed in the experiment. (b) For mean photon number $\bar{n}_a=0.05$ the theory prediction based on Eq.~\eqref{mean photon} also reasonably matches the amplitude of the peak-dip feature. (c) Additional poles are predicted if the $\mathsf{b}$ mode (associated with the array) is occupied. In each panel (a)--(c), significant deviations persist close to half-integer $\Phi_\text{ext}/\Phi_0$. (d) Fluxonium transition and harmonic mode frequencies, as specified by labels. Vertical dashed lines show alignment of resonances with corresponding poles in dispersive shifts and experimental features.}
\label{panel2}
\end{figure}

As one cause for deviations, we note that the fluxonium $0$-$1$ transition reaches a minimum frequency close to $300\,\text{MHz}$ at half-integer flux [see Fig.~\ref{panel2}(d)]. For a typical temperature of $T=20\,\text{mK}=0.42\,h\,\text{GHz}/k_{\!B} $,  thermal excitation of the $l\!=\!1$ level indeed becomes relevant.  (Thermal occupation of higher states $l > 1$ remains negligibly small.) We account for thermal excitation under the simplifying assumption that the measurement probes a statistical mixture of the lowest two fluxonium states with simple Boltzmann weights. In this case, the thermally averaged value of the dispersive shift of the resonator is given by
\be\label{Boltzman}
\langle\mu_{a;l}(n_a, n_b)\rangle=\frac{\mu_{a;0}(n_a, n_b)+e^{-\epsilon_{10}/k_{\!B} T}\mu_{a;1}(n_a, n_b)}{1+e^{-\epsilon_{10}/k_{\!B} T}},
\ee
which we use for the remainder of this subsection. We expect thermal effects only to be significant for the flux range $0.35 \alt \abs{\Phi_\text{ext}/\Phi_0} \le 0.5$; outside this range, $\epsilon_{10}$ exceeds $2\,\text{GHz}$ and thermal excitations of the fluxonium device should be negligible.

We now turn to the discussion of fourth-order corrections to the dispersive shifts.
In Fig.\ \ref{panel2}(a), (b) and (c),  
we compare  the same experimental data for the homodyne signal with the theoretical calculations  now including all fourth-order corrections and taking into account thermal averaging. Specifically, we calculate the homodyne voltage [Eq.\ \eqref{homodyne_fit}] using the same fit parameters as above and replace $\chi\to\langle \mu\rangle$. For easy reference of resonances, panel (d) shows the fluxonium transition frequencies as well as integer combinations of the harmonic frequencies $\omega_a$ and $\omega_b$. Relevant resonances are circled and labeled by roman numerals. Their alignment with the corresponding poles in (a)--(c) is indicated by vertical dashed lines. 

\begin{figure}
\includegraphics[width=0.65\columnwidth]{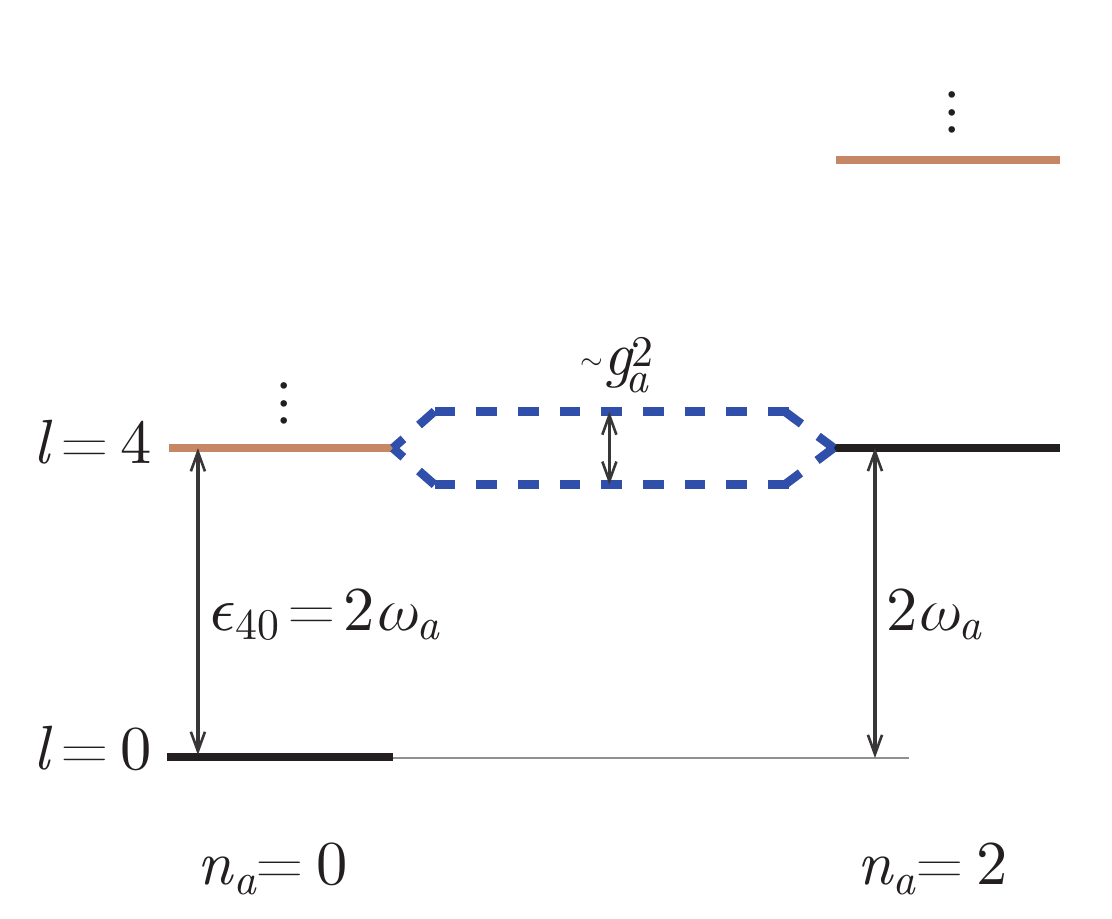}
\caption{(Color online)  Level diagram illustrating a two-photon vacuum Rabi splitting.   In this diagram, horizontal solid lines represent bare fluxonium levels which are coupled via a single photon mode. Here, the bare states with $n_a=0$, $l=4$ and $n_a=2$, $l=0$  are degenerate and coupled by an effective two-photon transition.  The resulting two-photon vacuum Rabi splitting is represented by the dashed lines.}   
\label{towphotonrabi}
\end{figure}

Figure \ref{panel2}(a) shows the thermally averaged homodyne voltage for the case of negligible harmonic mode excitations, $n_a=0, n_b=0$, and for the case of one initial excitation in the resonator mode, $n_a=1, n_b=0$. This choice of occupation numbers is motivated by the estimate of  $\langle n_a \rangle \approx 0.01$ as given in Ref.\ \onlinecite{manucharyan_fluxonium:_2009}.  In the flux region $0\le|\Phi_\text{ext}/\Phi_0|\alt0.35$, fourth-order contributions to the dispersive shift with initial state $n_a=0$  have no significant effect as compared to the second order results [Fig.\ \ref{dispersive_shifts}(b)].
The dispersive shift applicable to the $n_a=1$ state, however, does show an additional pole close to $|\Phi_\text{ext}/\Phi_0|\approx 0.15$ labeled by \{v\} in Fig.~\ref{panel2}(a). As indicated in panel (d), this pole occurs due to a resonance between two resonator photons, $2\omega_a$, and the 0-4 fluxonium transition, $\epsilon_{40}$ which we illustrate in Fig.\ \ref{towphotonrabi}.  Specifically, the pole in $\mu$ originates from a fourth-order term in the effective Hamiltonian, which is associated with the $V^+_aV^+_aV^-_aV^-_a$ path of the ladder diagram [Fig.\ \ref{fig3}]:
\be\label{pole1}
\sum_{l_1,l_3}\frac{g_{a;0l_1}g_{a;l_14}g_{a;4l_3}g_{a;l_30}}{(\omega_a+\epsilon_{0l_3} )(2\omega_a-\epsilon_{40})(\omega_a+\epsilon_{0l_1})} \mathsf{n}_a(\mathsf{n}_a-1) \ketbra{0}.
\ee
Note that this term vanishes for $n_a=0$ or $n_a=1$.  Hence, according to Eq.~\eqref{nonlinearshifts} it contributes to $\mu_{a;0}(n_a=1, n_b)$ which involves photon absorption $n_a=1\to2$  but not  to $\mu_{a;0}(n_a=0, n_b)$ where photon absorption occurs as $n_a=0\to1$. This fact is easily visible in the two theory curves shown in Fig.~\ref{panel2}(a). The flux position of the pole is in excellent agreement with a similar feature in the experimental data.  We  note that this resonance also manifests as the two-photon vacuum Rabi splitting illustrated in Fig.\ \ref{towphotonrabi}, with splitting size $2\sqrt{2}\sum_{l'}g_{a; 0l'}g_{a; l'4}/\Delta_{a; 0l'}$ (see Appendix \ref{twophotonrabi} for derivation).  As opposed to the usual vacuum Rabi splitting, the two-photon splitting is proportional to $g_{a}^2$ rather than $g_{a}$. As a direct consequence of the absence of strict selection rules for fluxonium, the summation of contributions from multiple intermediate states $l'$ can lead to a sizable splitting.

To compare the amplitude of the pole feature  with the experimental reflection data, we show the weighted average of the dispersive shifts for $n_a=0$  and $n_a=1$ in Fig.~\ref{panel2}(b). We parametrize the respective weights $P_0$ and $P_1=1-P_0$ (probabilities for the two initial resonator states) in terms of the mean photon number $\bar{n}_a=1-P_0$. The weighted average is given by
\begin{align}\label{mean photon}
 \bar{\mu}_{a}(\bar{n}_a)=&P_0 \langle\mu_{a;l}(0,0)\rangle  + (1-P_0)\langle \mu_{a;l}(1,0)\rangle. 
\end{align}
For a mean photon number of $\bar{n}_a=0.05$ (slightly higher than the value $0.01$ reported in Ref.~\onlinecite{manucharyan_fluxonium:_2009}), we find good agreement between theory and experiment for the amplitude of the resonance \{v\}.

In summary, for the flux region $0 < |\Phi_\text{ext}/\Phi_0| \alt 0.3 $ we find very good agreement between theoretical prediction and the experimental data for the dispersive shifts, including the positions and amplitudes of the two resonances \{v\} and \{vi\}. In the flux region closer to half-integer $\Phi_\text{ext}/\Phi_0$, however, agreement between experimental data and theory is weaker. As seen in Fig.~\ref{panel2}(a) and (b), the poles \{ii\} and \{iii\} due to resonances of $\epsilon_{40}$ and $\epsilon_{10}$ with $2\omega_a$ and $\omega_b-\omega_a$, respectively, do not quantitatively match the experimental data in this flux region, which are dominated by a pronounced minimum at $|\Phi_\text{ext}/\Phi_0|\approx0.38$.  The resonance features in the experimental data near $|\Phi_\text{ext}/\Phi_0|\approx0.325$ and $0.44$ are absent in the calculated dispersive shifts of panels (a) and (b). Vice versa,  the pole \{i\} predicted by theory has no correspondence in the experimental data.  We next discuss the effects of $\mathsf{b}$ mode occupations, which may give partial explanations for some of the mismatches. 

In panels (a) and (b) of Fig.\ \ref{panel2}, amplitudes predicted for the resonances \{ii\} and \{iii\} are dramatically smaller than features observed in the experiment at corresponding flux values. While thermal excitation of the $\mathsf{b}$ mode can be ruled out due to the large gap between the resonant frequency $\omega_b/2\pi=10.79\,\text{GHz}$ and the frequency $0.42\,\text{GHz}$ associated with a temperature of  $T\sim 20\,\text{mK}$, it is instructive to inspect the effects of nonequilibrium excitations of this mode. For this purpose, Fig.~\ref{panel2}(c) shows the homodyne voltage in the presence of one  $\mathsf{b}$ mode excitation,  as obtained from the dispersive shift $\langle\mu_{a;l}(0,n_b=1)\rangle$. As an important result of this excitation, the amplitude of resonance \{iii\} is amplified significantly. The enhancement of the resonance stems from four terms associated with the dual-mode ladder diagrams in Fig.\ \ref{fig456}(b) and (c).  As an example, we give the expression for one of them:
\be
\sum_{l_ 1,l_3}\frac{g_{a; 0l_1}g_{b; l_11}g_{b; 1 l_3}g_{a; l_30} (\mathsf{n}_a+1)\mathsf{n}_b}{(-\omega_a+\epsilon_{0l_3} )(\omega_b-\omega_a-\epsilon_{10})(-\omega_a+\epsilon_{0l_1})} \ketbra{0},
\ee
which corresponds to the  $V^-_aV^+_bV^-_bV^+_a$ path. The other three terms arise from the paths $V^+_bV^-_aV^+_aV^-_b$, $V^+_b V^-_a V^-_b V^+_a$, and  $V^-_a V^+_b V^+_a V^-_b$. They lead to the contributions to the effective Hamiltonian of the form $(\mathsf{n}_a+1)\mathsf{n}_b \ketbra{0}$ and involve the  denominator $(\omega_b-\omega_a-\epsilon_{10})$.  All of these terms indeed vanish  for $n_b=0$ and hence do not contribute to the previous results in panels (a) and (b). It is thus possible that nonequilibrium $\mathsf{b}$ mode excitations are partly responsible for the deviations between theory and experiment in the half-integer flux region.

By a similar mechanism, $\mathsf{b}$ mode excitations also lead to an additional predicted resonance \{iv\} at a flux position fairly close to the resonance feature in the experimental data at  $|\Phi_\text{ext}/\Phi_0|\approx 0.325$. According to theory, this resonance occurs whenever the array mode excitation $\omega_b$ matches the fluxonium 1-3 transition $\epsilon_{31}$.   The corresponding term in the effective Hamiltonian is
\be
\sum_{l_2, l_3}\frac{2g_{a; 13}g_{a; 3l_2}g_{b; l_2l_3}g_{b; l_31}}{(-\omega_a+\epsilon_{1l_3} )\epsilon_{1l_2}(\omega_b-\epsilon_{31})} (\mathsf{n}_a+1)\mathsf{n}_b \ketbra{1},
\ee
associated with the $V^+_bV^-_bV^+_aV^-_a$ and $V^+_aV^-_aV^+_bV^-_b$ paths in the dual-mode ladder diagram of Fig.\ \ref{fig456}(a). We have verified that  a further increase in the $\mathsf{b}$ photon number beyond $n_b=1$ [shown in Fig.~\ref{panel2}(c)] does not improve the fit.

In summary, we find that nonequilibrium array-mode excitations may produce significant changes in the dispersive shifts, some of which may point to resonance features observed in the experiment.  Without a detailed understanding of the underlying nonequilibrium distribution and its dependence on magnetic flux, however, it is difficult to assess whether this explanation could ultimately give a quantitative match or whether additional array degrees of freedom, breakdown of perturbation theory, or dynamical effects under continuous driving are responsible for the observed deviations.

\begin{figure*}
\includegraphics[width=0.85\textwidth]{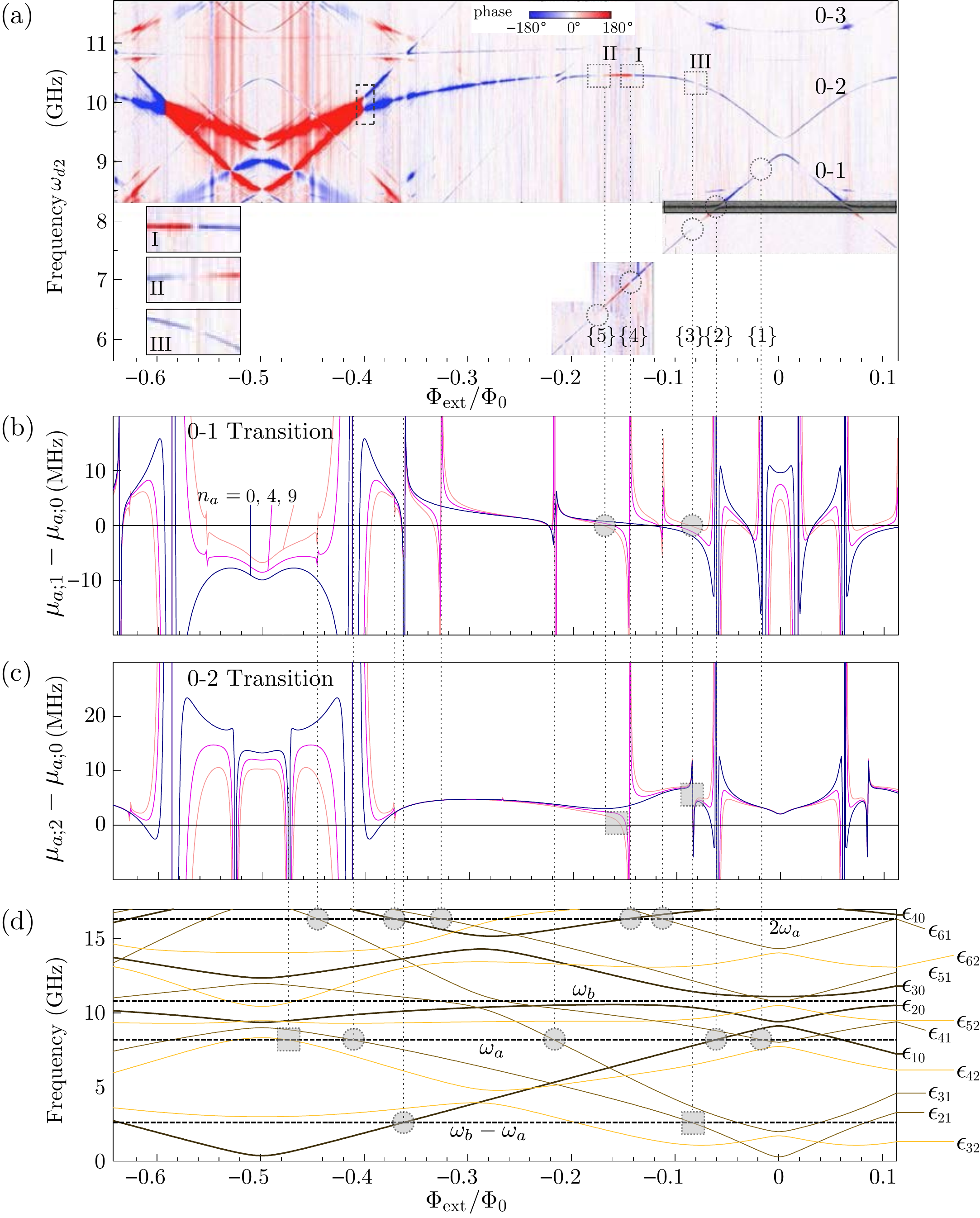}
\caption{(Color online)   (a) Two-tone spectrum. The vertical axis represents the frequency of the sweeping tone, $\omega_{d2}$, and the horizontal axis represents the external magnetic flux. The color-scale represents the phase difference between the ground state and the state $l$ excited by the sweeping tone, namely $\theta_l-\theta_0$, where red represents positive value and blue negative. The three types of color changing, I, II and III are illustrated by magnified insets. The color  change marked by the dashed box corresponds to a spurious phase wrapping (see text).    (b) Plot of the nonlinear dispersive shifts $\mu_{a;1}(n_a, n_b)-\mu_{a;0}(n_a, n_b)$, where $n_b=0$ for all three curves, representing shifts with photon numbers $n_a=0,4,9$ respectively. (c) Plot of the nonlinear dispersive shifts $\mu_{a;2}(n_a, n_b)-\mu_{a;0}(n_a, n_b)$ with photon numbers $n_a$ and $n_b$ as in (b).  (d) Plot of qudit transition frequencies and harmonic mode frequencies. Horizontal dashed lines show harmonic frequencies. Solid curves represent qudit  frequency differences and coloring distinguishes transitions starting from $l$$=$$0$, $1$, and $2$. Circles and squares mark zero points, resonances (and corresponding poles) and their alignment with experimental data.} 
\label{artificialspectrum}
\end{figure*}

\subsection{Two-tone spectroscopy}
As explained in subsection \ref{sec:measurements}, two-tone spectroscopy probes the change in the dispersive resonator shift due to induced Rabi oscillations between the fluxonium ground state and another fluxonium state (level $l$). When detected via the change in the phase of the reflected homodyne signal, the relevant observable is given by $\theta_{l0} \equiv \theta_l-\theta_0$ as obtained from Eq.~\eqref{arctanfunction}. This phase difference is  encoded by the color scale in Fig.\ \ref{artificialspectrum}(a), which shows experimental data from Refs.\ \onlinecite{manucharyan_fluxonium:_2009, phase_slip_2012,Vlad-thesis}. The observed transition lines correspond to the  fluxonium 0--1, 0-2, and 0-3 transitions and to an array-mode with frequency $\omega_b=10.79\,\text{GHz}$.  

Interestingly, the spectroscopy data does not only reveal the frequencies of relevant transitions but also contains several distinct phase changes along these transition lines. In particular, the phase changes observed in the experiment can be classified into four types. We use the 0-2 transition line for illustrating these types and refer to the labels and magnified insets in Fig.~\ref{artificialspectrum}(a): type I is an \textbf{abrupt red-blue change}, corresponding to a sudden jump from positive to negative phase difference $\theta_{l0}$; type II is a \textbf{gradual blue-white-red change}, corresponding to a continuous change of the phase difference from negative to positive values (or vice versa); type III is a \textbf{gradual blue-white-blue change}, corresponding to a the phase difference approaching zero and recovering without changing its sign. In principle, \textbf{red-white-red color changes} may also occur (type IV) but are not realized in this instance of data.

In the following, we show that fourth-order contributions to the dispersive shifts explain these phase changes.
The  relation between the dispersive shift and the phase difference is given by
\begin{align}
\theta_{l0}=&2 \ \text{arctan}\bigg( \frac{2 Q}{\omega_a}[\omega_{d1}-\omega_a-\mu_{a;l}]\bigg)\nonumber\\
&-2 \ \text{arctan}\bigg( \frac{2 Q}{\omega_a}[\omega_{d1}-\omega_a-\mu_{a;0}]\bigg).
\label{phasedifference}
\end{align}
This expression follows from Eq.\ \eqref{arctanfunction} when setting $\omega_d=\omega_{d1}$. Our discussion of the different types of phase changes will be based on identifying magnetic flux values where the phase difference vanishes,  $\theta_{l0}=0$ (type II and III), and flux values where the dispersive shifts have poles such that the phase difference may jump from $\pi$ to $-\pi$ (type I). The first condition is satisfied whenever $\mu_{a;l}-\mu_{a;0}=0$, the second condition whenever appropriate resonances between fluxonium transitions and harmonic modes occur.
We note that in the linear regime of the two $\arctan$ functions, the phase difference is simply given by 
\be\label{phasesign}
\theta_{l0} = -4Q(\mu_{a;l}-\mu_{a;0})/\omega_a.
\ee
 Thus, in the linear regime, the two-tone spectroscopy serves as a direct probe of the dispersive shifts.

For comparison with the experimental data, we calculate the nonlinear shifts for the fluxonium transitions $l$$=$$0$$\to$$1$ and $l$$=$$0$$\to$$2$, and show the corresponding differences $\mu_{a;1}-\mu_{a;0}$ and $\mu_{a;2}-\mu_{a;0}$ in Fig.\ \ref{artificialspectrum}(b) and (c), respectively. In each case, we consider dispersive shifts for different photon numbers $n_a=0, 4, 9$ to illustrate the dependence of the detected phase response on the power of the microwave probe tone.  We use vertical dashed lines with labels to emphasize the alignment of poles and zero points  with the corresponding phase changes. According to Eq.\ \eqref{phasesign}, the color-encoded sign of the phase difference in panel (a) thus corresponds  to the sign of $-(\mu_{a;l}-\mu_{a;0})$ in panels (b) and (c), respectively. We compare theoretical and experimental results for the 0-1 and 0-2 transitions in the flux regions $-0.22\Phi_0 \le \Phi_\text{ext} \le 0$ and $-0.5\Phi_0 \le \Phi_\text{ext} \le 0$ where experimental data is available.\cite{manucharyan_fluxonium:_2009, phase_slip_2012, Vlad-thesis} The following discussion is organized according to the three observed types of phase changes.

\subsubsection{Type I -- abrupt red-blue phase change}
The only type I phase change in Fig.\ \ref{artificialspectrum} is marked by the vertical line with label \{4\}, and occurs due to a pole appearing in both $\mu_{a;1}-\mu_{a;0}$ and $\mu_{a;2}-\mu_{a;0}$. This pole originates from the fourth order contribution to $\mu_{a;0}$ given in Eq.\ \eqref{pole1} and corresponds to a resonance between two resonator photons and the fluxonium 0-4 transition, $2\omega_a=\epsilon_{40}$, see Fig.\ \ref{artificialspectrum}(d).  (This is the same resonance that  also gives rise to the two-photon vacuum Rabi splitting illustrated in Fig.\ \ref{towphotonrabi}.)  As noted before, this pole only occurs when $n_a \ge 1$. As a result,  the $n_a=0$ curves in panels (b) and (c) do not exhibit this pole, and the pole becomes more pronounced as the photon number increases. We thus conclude that spectroscopy of the fluxonium transitions, in this case, also reveals information about photon population in the resonator.

Two clarifications are in order. First, the absence of a type I phase change in the 0-2 transition due to the pole in $\mu_{a;2}-\mu_{a;0}$ marked by the vertical line \{2\} falls outside our current discussion: at this flux value, the 0-1 transition is resonant with $\omega_a$ and the discussion of the phase shift would need to include the hybridization of photon and fluxonium excitation. Second, we note that the abrupt phase change close to $\Phi_\text{ext}=-0.4\Phi_0$, marked by a  dashed box in panel (a), is not associated with a pole and hence \emph{not} of type I. Since phase changes in the experiment are defined in the interval $[-\pi,\pi]$, such additional phase jumps may simply occur when dispersive shifts become sufficiently large so that the magnitude of $\theta_{l0}$ exceeds $\pi$. These phase discontinuities do not involve sign changes in the dispersive shifts $\mu_{a;l}-\mu_{a;0}$. Indeed, the occurrence of such a phase discontinuity close to half-integer $\Phi_\text{ext}/\Phi_0$ is consistent with the large magnitude of the dispersive shifts $\mu_{a;2}-\mu_{a;0}$ predicted by theory in this region. The theoretically predicted pole corresponding to the resonance between $\epsilon_{21}$ and $\omega_a$ is slightly on the left of the dashed box.  Close to $\Phi_\text{ext}=-0.41\Phi_0$, experimental evidence for the avoided crossing of the two spectral lines (the 0-2 transition and the 0-1 transition  shifted by the photon frequency $\omega_a$)  with better resolution can be found in Ref.\ \onlinecite{phase_slip_2012}.

\subsubsection{Type II -- gradual blue-white-red change}
Gradual phase changes of type II are present in both the 0-1 and the 0-2 transitions and are marked by the vertical line with label \{5\}. In both cases, the dispersive shift $\mu_{a;l}-\mu_{a;0}$ ($l=1,2$) smoothly crosses through zero so that the phase change is negative for flux values $<-0.18\Phi_0$, reaches zero, and then assumes positive values as $\mu_{a;l}-\mu_{a;0}$ approaches the pole at position \{4\}. As shown in panels (b) and (c), the precise zero-point crossing is, in fact, photon-number dependent. We note that the alignment between the predicted crossings for photon numbers as large as $n_a=9$ is not perfect. Quite likely, this can be attributed to the breakdown of perturbation theory in the immediate vicinity of poles and hence especially applies to the zero-crossing for $\mu_{a;2}-\mu_{a;0}$. As for the absence of pole \{4\} for $n_a=0$ discussed above, we also expect the crossing \{5\} to disappear if no resonator photons are present.

An additional phase change of type II is present only in the fluxonium 0-1 transition and is marked  by line \{3\}  and can be interpreted in a similar manner. Again, we find alignment of the zero crossing with the experimental feature only for rather large photon numbers, see the  $``n_a=9"$ curve in panel (b). An experimental study of the power dependence of this phase change could help shed more light on the quantitative comparison with theory.

\subsubsection{Type III: gradual blue-white-blue color change}
Type III phase changes occur for both the 0-1 and 0-2 transitions, and instances are marked by the vertical dashed lines labeled by \{1\} and \{3\} respectively. In both cases, we observe alignment with poles occurring in the corresponding dispersive shifts. A definite prediction for type III phase changes, however, appears difficult based on the perturbative results. In general, perturbation theory will break down at the position of the pole and will remain unreliable in a certain flux window in its vicinity. As a result, predictions in this case must likely be based on non-perturbative methods and may possibly also have to take into account the dynamical aspect of the two-tone measurement (which are beyond the scope of this paper). Qualitatively, the type III phase changes are at least plausible given that the dispersive shifts $\mu_{a;l}-\mu_{a;0}$ are predominantly positive in the direct vicinity of both poles.


\section{Conclusion\label{conclusions}}
In summary, we have presented a systematic treatment of fourth-order corrections to the dispersive regime of circuit QED. Our results, developed in Section II, are valid for a generic system consisting of a multi-level qudit capacitively coupled to one or several harmonic modes, and hence apply to a wide class of circuit QED systems. Our treatment, in particular, enables the description of dispersive shifts in systems lacking simplifying selection rules.

We have applied our results to the concrete case of the fluxonium device as realized in recent experiments.\cite{manucharyan_fluxonium:_2009,phase_slip_2012, Vlad-thesis}
Using numerical diagonalization, we have obtained the relevant charge matrix elements which confirm the lack of selection rules, and have incorporated them in the perturbative treatment of the dispersive regime, including corrections up to fourth order. The calculated dispersive shifts allow us to compare theoretical predictions with experimental data for homodyne reflection measurements and two-tone spectroscopy from Refs.~\onlinecite{manucharyan_fluxonium:_2009,  phase_slip_2012,Vlad-thesis}.

The absence of selection rules is an important mechanism for producing sizable dispersive shifts, even if the transition of interest is far detuned from the resonator used for readout. For the fluxonium system studied in Section III, our calculations show that dispersive shifts can indeed be as large as $10$ MHz even when the corresponding 0-1 fluxonium transition is detuned by almost $8$\,GHz from the resonator. The lack of selection rules enables a multitude of virtual transitions to contribute to the dispersive shifts. Especially if such higher transition frequencies match photon resonance conditions, dispersive shifts can be surprisingly large.
We also note that the magnitudes of matrix elements are tunable with external magnetic flux, which in turn leads to the tunability of dispersive shifts. 

Away from half-integer $\Phi_\text{ext}/\Phi_0$, we find good quantitative agreement for the homodyne reflection data with our theory prediction. This agreement also includes a resonance feature in the data which previously remained unexplained. The presence of this resonance indicates a small probability of a photon occupying the resonator, so that its amplitude may be used for extracting the mean photon number. Close to half-integer $\Phi_\text{ext}/\Phi_0$, even though we find tentative agreement between the flux positions of several resonance features in experimental data and theory, our calculation does not give a  quantitative match. We note that the flux position coincidence of resonances points to nonequilibrium array-mode excitations of unknown origin. 

Spectroscopy data of fluxonium samples\cite{manucharyan_fluxonium:_2009,  phase_slip_2012,Vlad-thesis} show unusual phase changes along the transition lines. We have identified three different types of phase changes, according to abrupt and gradual variations of the phase with or without sign change. Our calculations show that these phase changes are closely related to poles and zero points in the dispersive shifts and that their occurrence may sensitively depend on photon numbers in the resonator. Experimental studies of the power dependence of spectroscopy measurements are an interesting subject for future study and may shed additional light on the origin of quantitative deviations between experiment and theory. Our results for spectroscopic phase changes show generally good agreement for the flux positions of such resonances. The prediction of the specific type of the phase change remains challenging since perturbative calculations break down at the positions where resonances occur. Nonperturbative calculations, and taking into account the dynamics of the measurement protocol in a Master equation description, may be necessary to obtain such type of predictions and to improve quantitative agreement.  An additional source of quantitative deviations may lie in the presence of additional array modes not included in our description. The spectroscopy data indeed shows additional levels, especially close to $|\Phi_\text{ext}/\Phi_0|\simeq \tfrac{1}{2}$, which warrant further investigation.

For both types of experiments,  we have identified a two-photon resonance near $\pm 0.14\Phi_\text{ext}/\Phi_0$ which manifests in the dispersive shifts in fourth order of perturbation theory and which should also lead to a two-photon vacuum Rabi splitting.  Experimental verification would involve tuning the drive frequency $\omega_d$ close to the $\epsilon_{40}$ transition and directly observing the level splitting.  (In previous experiments the $\epsilon_{40}$ transition was outside the measured frequency range.)  Further experimental verification could be achieved by detecting the correlated emission of photon pairs under vacuum Rabi oscillation.

The accumulation of contributions to dispersive shifts in the absence of selection rules does not only affect ac-Stark shifts but can, similarly, lead to surprisingly large self-Kerr and cross-Kerr coefficients in fourth order which are tunable with magnetic flux.  Making photon-photon interaction terms large while keeping the fundamental qudit transition off resonance, is particularly appealing for  circuit QED lattices which have been discussed as quantum simulator architecture\cite{hartmann_quantum_2008,houck2012} and for which the first experimental realizations are now underway.\cite{underwood2012}

\begin{acknowledgments}
We  thank  Michel Devoret and Leonid Glazman for numerous insightful discussions. We further acknowledge Archana Kamal, Nicholas Masluk, James Sauls, Anupam Garg, Joshua Dempster and Andy Li for valuable discussions. We thank all authors of Refs.\ \onlinecite{manucharyan_fluxonium:_2009,  phase_slip_2012} for providing the previously unpublished data shown in Figs.\ \ref{panel2} and \ref{artificialspectrum}. Our research was supported by the NSF under Grant PHY-1055993.
\end{acknowledgments}

\appendix
\section{Adiabatic elimination}\label{adiabatic}
We consider a quantum mechanical system described by a Hilbert space $\mathcal{H}$, which is composed of several subspaces, $\mathcal{H}=\mathcal{H}_1\oplus\mathcal{H}_2\oplus\cdots$ For our following treatment, it is convenient to introduce the projectors 
\be
P_\alpha= \sum_{j} \ketbra{\alpha,j}
\ee
onto these subspaces.
Here, the index $j$ enumerates the eigenstates within each subspace $\mathcal{H}_\alpha$. The projectors satisfy the completeness condition
\be
\sum_\alpha P_\alpha= \openone.
\ee

The goal of adiabatic elimination is to construct a Hamiltonian $H'$ with the same eigenenergies as the original Hamiltonian but without coupling between the specified subspaces. The elimination is facilitated by a unitary transformation,
\be
H'=e^{iS}He^{-iS},
\ee
where the generator $ {S}$ is chosen such that
\be\label{condition1}
P_\alpha H'P_\beta = P_\alpha e^{iS}He^{-iS}P_\beta=0 \qquad (\alpha \neq \beta),
\ee
which eliminates all coupling between subspaces. Condition \eqref{condition1} only determines the off-diagonal elements of $H'$. To remove the remaining ambiguity, one imposes the constraint that the diagonal elements of the  generator $ {S}$  be zero,
\be\label{condition2}
P_\alpha S P_\alpha =0.
\ee
Conditions \eqref{condition1} and \eqref{condition2} then uniquely determine $ {S}$ and  $H'$.

To construct $S$ and $H'$ perturbatively and order by order, we decompose $S$ into a series,
\be
S = \sum_{n=1}^{\infty}{\lambda}^n S_n,
\ee
where $\lambda$ is an auxiliary parameter for order counting. 
Using Hadamard's lemma, we rewrite
\begin{align}
H' =& H + [iS,H]+\frac{1}{2!}[iS, [iS,H]]+\cdots 
=\sum_{n=0}^{\infty}\frac{1}{n!}[iS,H]_n.
\label{baker}
\end{align}
Decomposing $H'$ in a similar series
\be\label{Hexpansion}
H'=H_0+ \sum_{n=1}^{\infty}{\lambda}^nH'_n,
\ee
substituting into Eq.\ (\ref{baker}) and consistently collecting terms up to ${\lambda}^2$,  one obtains
\begin{align}\label{uptosecondorder}
\nonumber H' =&H+[iS,H]+\frac{1}{2} [iS,[iS,H]]+ \mathcal{O}(\lambda^3)\\
\nonumber =& H_0+\{\lambda V + [i\lambda S_1,H_0]\} + \{ [i\lambda^2 S_2, H_0] + [i\lambda S_1, \lambda V]\\
&+ \frac{1}{2} [i\lambda S_1, [i\lambda S_1,H_0]]\} +\mathcal{O}(\lambda^3).
\end{align}
After imposing conditions \eqref{condition1} and \eqref{condition2}, the resulting expression for $H'$ up to (and including) second order is given by\cite{cohen-tannoudji_atom-photon_1998} 
\be\label{Heff2}
H'=H_0+V^{\text{diag}} + \frac{1}{2} \sum_\alpha P_\alpha [iS_1, V^{\text{nd}}] P_\alpha + \mathcal{O}(\lambda^3).
\ee
Here, the diagonal components of $V$ are $V^\text{diag}=\sum_\alpha P_\alpha V P_\alpha$, and the off-diagonal elements $V^\text{nd}=V-V^\text{diag}$. To first order, the generator is obtained as 
\begin{align}\label{S1gen}
P_\alpha i\,S_1 P_\beta= \frac{P_\alpha V P_\beta}{E_\alpha- E_\beta}.
\end{align}

We then apply Eq.\ \eqref{Heff2} to a system composed of a multi-level qudit coupled to harmonic modes (such as the ones realized by a resonator). The non-interacting and interaction contributions to the Hamiltonian, $H_0$ and $ {V}$, are given by Eqs.\ \eqref{Hamiltonian1} and  \eqref{Hamiltonian2}. By our choice, the individual subspaces are one-dimensional each and spanned by one of the eigenstates of $H_0$, i.e., $\ket{\vc{n}l}_0$. The integer components of $\vc{n}=(n_j)$ denote the photon numbers of the harmonic modes (indexed by $j$),  and $l=0,1,\ldots$ enumerates the qudit levels.

Since $V$ involves an increase or decrease in photon number, its diagonal part vanishes, $V^{\text{diag}}=0$, and thus one finds $V=V^{\text{nd}}$.
From Eq.\ \eqref{S1gen}, we obtain the explicit form of the first-order generator
\begin{align}\label{generators1}
S_1 = -i \sum_j \sum_{ll'}\bigg[\frac{g_{j;ll'}a_j}{\epsilon_{ll'} - \omega_j} -\frac{g_{j;l'l}a^{\dag}_j}{\epsilon_{l'l} - \omega_j}\bigg] \ket{l}_0{_0\bra{l'}},
\end{align}
where $\epsilon_{l'l}\equiv\epsilon_{l'}-\epsilon_l$ is a transition energy.
Finally, Eq.\ \eqref{Heff2} yields the following second-order expression for $H'$: 
\begin{align}\label{activemethod}
H' =&H_0 + \sum_{ll' }\sum_j[(\chi_{j;ll'} - \chi_{j;l'l}) a^\dag_j a_j + \chi_{j;ll'}] \ket{l}_0{_0\bra{l}}.
\end{align}
Here, we have abbreviated the partial dispersive shifts by
\be\label{2ndshifts}
\chi_{j;ll'} = \frac{|g_{j;ll'}|^2}{\epsilon_{ll'}-\omega_j} \equiv \frac{|g_{j;ll'}|^2}{\Delta_{j;ll'}}
\ee
By construction, $H'$ [Eq.\ \eqref{activemethod}] is turned into the effective Hamiltonian $H_\text{eff}$ in Eq.\ \eqref{second} by replacing bare operators by dressed ones, $a_j\to\mathsf{a}_j$ etc.  The expression for the second-order shifts in Eq.\ \eqref{2ndshifts} agrees with the one  derived in the main text [Eq.\ \eqref{coeffs1}].

\section{Numerical diagonalization of the fluxonium Hamiltonian \label{app:diag}}
Numerical diagonalization of the fluxonium Hamiltonian $H_\text{f}$, Eq.\ \eqref{Hf} can be achieved by using the harmonic oscillator basis $\{\ket{m}\}$ which diagonalizes $H_\text{f}|_{E_J=0}$. The oscillator states $\ket{m}$ are generally \emph{not} good approximations to the fluxonium eigenstates. Thus, proper care must be taken to ensure convergence of results with respect to the cutoff in $m$. Considering $\varphi$ as the position coordinate, $\varphi_0=(8E_C/E_L)^{1/4}$ plays the role of an oscillator length. The corresponding oscillator eigenenergies are given by $\epsilon_m=\hbar\omega_p(c^\dag c+1/2)$, where $c^\dag$ and $c$ are the usual raising and lowering operators  and $\omega_p=\sqrt{8E_LE_C}/\hbar$ denotes the oscillator frequency. 

In this basis, the Hamiltonian matrix has the form
\begin{align*}
H_{mm'}=\delta_{mm'}\epsilon_m &-E_J \cos(2\pi\Phi/\Phi_0) \boket{m}{\cos\varphi}{m'}\\
&-E_J \sin(2\pi\Phi/\Phi_0) \boket{m}{\sin\varphi}{m'}.
\end{align*}
The sine and cosine matrix elements can be expressed in terms of generalized Laguerre polynomials as\cite{jeffrey_table_2007}
\begin{align}\label{cosexpansion}
&\boket{m}{\cos\varphi}{m+2p}\\\nonumber
&=(-2)^{-p}\sqrt{\frac{m!}{(m+2p)!}}\varphi_0^{2p}e^{-\varphi_0^2/4}\text{L}_m^{2p}(\varphi_0^2/2)\\\nonumber
&\boket{m}{\sin\varphi}{m+2p+1}\\\nonumber
&=(-2)^{-p}\sqrt{\frac{m!}{2(m+2p+1)!}}\varphi_0^{2p+1}e^{-\varphi_0^2/4}\text{L}_m^{2p+1}(\varphi_0^2/2),
\end{align}
where $m,p\in\NN$. 

Alternatively, one may employ the direct evaluation of matrix-valued exponentials. Specifically, the cosine term may be re-expressed via
\begin{align*}
\cos(\varphi-\varphi_\text{ext})=&\frac{1}{2}e^{i\varphi}e^{-i\varphi_\text{ext}}+\text{H.c.}\\
=&\frac{1}{2}\exp\left[\tfrac{i\varphi_0}{\sqrt{2}}(c+c^\dag)\right]e^{i\varphi_\text{ext}}+\text{H.c.},
\end{align*}
Algorithms for matrix exponentiation are standard in commercial as well as non-commercial software and libraries, e.g., Mathematica: \texttt{MatrixExp[]}, Matlab: \texttt{expm()}, SciPy: \texttt{expm()} and \texttt{expm2()}.  

Once the Hamiltonian matrix is diagonalized,   $H_\text{f}= \sum_l \epsilon_l \ket{l}\bra{l}$, one obtains the charge matrix elements in the diagonalized basis, $\boket{l}{\mathsf{N}}{l'}$, as follows.  The charge operator is rewritten in the oscillator basis via $\mathsf{N}=-\frac{i}{\sqrt{2} \varphi_0} (c-c^\dag)$ and thus one finds
\begin{align}
\nonumber \boket{m}{\mathsf{N}}{m'}=&-\frac{i}{\sqrt{2}\varphi_0}(\sqrt{m'}\delta_{m,m'-1}-\sqrt{m}\delta_{m,m'+1}).
\end{align}
Finally, one obtains 
\be
\boket{l}{\mathsf{N}}{l'}=\sum_{mm'}\bket{l}{m}\boket{m}{\mathsf{N}}{m'}\bket{m'}{l'},
\ee
where $\bket{l}{m}$ and $\bket{m'}{l'}$ are eigenvector amplitudes in the harmonic oscillator basis.

\vspace*{3mm}
\section{Two-photon vacuum Rabi splitting\label{twophotonrabi}}
Consider a multi-level qudit coupled to a single resonator mode $a$ in a configuration where the bare states $\ket{l_1, n_a}_0$ and $\ket{l_2, n_a+2}_0$ are degenerate. The leading-order coupling between them is through a two-photon process involving the intermediate states  $\ket{l', n_a+1}_0$, with $l'=0,1,\ldots$   We derive the effective two-photon coupling  by adiabatic elimination of  the intermediate states $\ket{l', n_a+1}_0$ and obtain an effective Hamiltonian $H'_\alpha$ for the subspace spanned by the two states $\ket{l_1, n_a}$ and $\ket{l_2, n_a+2}$. By applying Eqs.\ \eqref{Heff2} and \eqref{generators1} and using the projector $P_\alpha$ for the relevant subspace, we find\begin{align}
H'_{\alpha}=& \sum_{l'}\frac{g_{a; l_2 l'}g_{a; l'l_1}}{\Delta_{a; l_2 l'}}\mathsf{a}^\dag\mathsf{a}^\dag\ket{l_2}\bra{l_1}+\text{H.c.}
\end{align} 
Here, the detuning is $\Delta_{a; l_2 l'}=\epsilon_{l_2 l'}-\omega_a$ as defined previously.
Note that the effective coupling indeed involves the emission or absorption of two photons. Diagonalization of $H'_\alpha$ yields the splitting
\be
\sum_{l'}\frac{2\sqrt{(n_a+1)(n_a+2)}g_{a; l_2 l'}g_{a; l'l_1}}{\Delta_{a; l_2 l'}}
\ee
for initial photon number $n_a$. With $n_a=0$, $l_1=4$ and $l_2=0$, we recover the expression for the two-photon vacuum Rabi splitting presented in the main text.


\begin{thebibliography}{45}
\bibitem[{exp()}]{exp_parameters}
\bibinfo{note}{Experimental parameters: $E_J/h = 8.91\,\text{GHz}; E_C/h =
  2.48\,\text{GHz}; E_L/h = 0.53\,\text{GHz}; \omega_a/2\pi=8.18\,\text{GHz};
  \omega_b/2\pi=10.78\,\text{GHz}; g_a/h=0.18\,\text{GHz}; g_b/h=0.39\,\text{GHz}$.   All the parameters are fit from the spectrum in fig.\ \ref{artificialspectrum}(a), by diagonalizing the coupled cQED system described by Eq.\ \eqref{Hamiltonian1} and Eq.\ \eqref{Hamiltonian2}  (including the fluxonium qudit and the two harmonic modes)  }.
\end{thebibliography}
\end{document}